\documentclass[twocolumn,prl,aps,superscriptaddress,longbibliography]{revtex4-2}
\usepackage{amsmath}
\usepackage{amssymb}
\usepackage{amsfonts}
\usepackage[dvips]{graphicx}
\usepackage{subfigure}
\usepackage{dcolumn}
\usepackage{txfonts}
\usepackage{bm}
\usepackage{makeidx}
\usepackage{color}
\usepackage{mathtools}
\usepackage{threeparttable}
\usepackage[colorlinks,linkcolor=blue,anchorcolor=blue,citecolor=blue,urlcolor=blue]{hyperref}
\usepackage{lipsum}
\usepackage{braket}

\begin{document}

\title{Rayleigh Diffraction Threshold Controls Nonlinear Photon Correlations in Atomic Arrays}
\author{Zai-Cheng Xiao}

\author{Li-Ping Yang}
\email{lipingyang87@gmail.com}
\affiliation{Center for Quantum Sciences and School of Physics, Northeast Normal University, Changchun 130024, China}

\begin{abstract}
Two-dimensional atomic arrays provide versatile free-space quantum optical interfaces by coupling photons to collective lattice modes. Here we show that the Rayleigh anomaly provides a sharp control mechanism for both linear and nonlinear photon scattering from a single atomic layer. As the lattice spacing crosses the diffraction threshold, newly opened radiative channels rapidly broaden the dominant collective modes, converting specular reflection into diffuse off-axis scattering. Below threshold, the scattered photons exhibit pronounced opposite-momentum correlations and interference fringes, which are strongly reorganized and largely disappear above it. Remarkably, an exact cancellation between the disconnected and part of the connected two-photon amplitudes leaves the reflected coincidence signal governed entirely by the two-excitation response. Its integrated scattering weight collapses by more than three orders of magnitude across the threshold. These results establish radiative diffraction thresholds as a control principle for quantum nonlinear optics in single-layer atomic arrays.
\end{abstract}
\maketitle

\textit{Introduction}---Ordered arrays of quantum emitters provide a versatile platform for engineering free-space light--matter interactions through deterministic atom-by-atom assembly and single-atom-resolved control and spectroscopy~\cite{Barredo2016Atom,Endres2016Atom,Hofer2025SingleAtom}. In two-dimensional lattices, photon-mediated resonant dipole--dipole interaction (RDDI) and lattice interference give rise to collective radiative eigenmodes with cooperative frequency shifts, modified linewidths, and directional emission~\cite{Lehmberg1970Radiation,Jenkins2012Controlled,Bettles2015Cooperative,Shahmoon2017Cooperative,AsenjoGarcia2017Selective}. These collective modes enable enhanced optical cross sections~\cite{Bettles2016Enhanced}, cooperative resonances and near-unity reflection~\cite{Shahmoon2017Cooperative,Facchinetti2018Interaction,Rui2020Subradiant,Srakaew2023Subwavelength}, and long-lived subradiant storage~\cite{Facchinetti2016Storing,AsenjoGarcia2017Selective,Ballantine2021quantum}. More broadly, atomic arrays realize quantum metasurfaces in which the spatial and quantum properties of light can be controlled within an atomically thin geometry~\cite{Perczel2017Topological,Bekenstein2020Quantum,Wang2022NonHermitian}. 

Beyond linear optics, emitter saturation and interatomic interactions generate many-body correlations and strong photonic nonlinearities~\cite{Bettles2020Quantum,Williamson2020Superatom,Parmee2021Bistable,MorenoCardoner2021Quantum,Zhang2022PhotonPhoton,Rusconi2021Photonic}. Rydberg interactions can turn two-dimensional arrays into strongly nonlinear quantum mirrors~\cite{MorenoCardoner2021Quantum,Zhang2022PhotonPhoton}. Coupled layers can act as nonlinear metasurfaces that convert classical inputs into correlated light~\cite{Pedersen2023Metasurfaces}. Even simple two-level arrays exhibit nonlinear few-photon scattering through saturation and multi-excitation dynamics~\cite{Williamson2020Superatom,Parmee2021Bistable,Rusconi2021Photonic}. Related two-dimensional waveguide-QED systems likewise support strongly interacting photon states~\cite{Tecer2024Strongly}. In addition to optical-depth bounds on many-body decay~\cite{Rusconi2026OpticalDepth}, radiative coupling in periodic emitter arrays is constrained by reciprocal-lattice momentum. Collective Bloch modes are labeled by transverse quasimomentum, and their radiative decay depends on which reciprocal-lattice-shifted momentum components lie inside the light cone~\cite{AsenjoGarcia2017Selective,Facchinetti2018Interaction,Ballantine2020Subradiance}. Increasing the lattice spacing can therefore open additional radiative diffraction channels~\cite{AsenjoGarcia2017Selective}. Although the opening of additional diffraction orders and the associated enhancement of collective radiative linewidths have been established~\cite{Shahmoon2017Cooperative}, their impact on frequency-optimized specular reflection and momentum-resolved two-photon correlations across the diffraction threshold has not been systematically investigated.

\begin{figure}
\includegraphics[width=8.5cm]{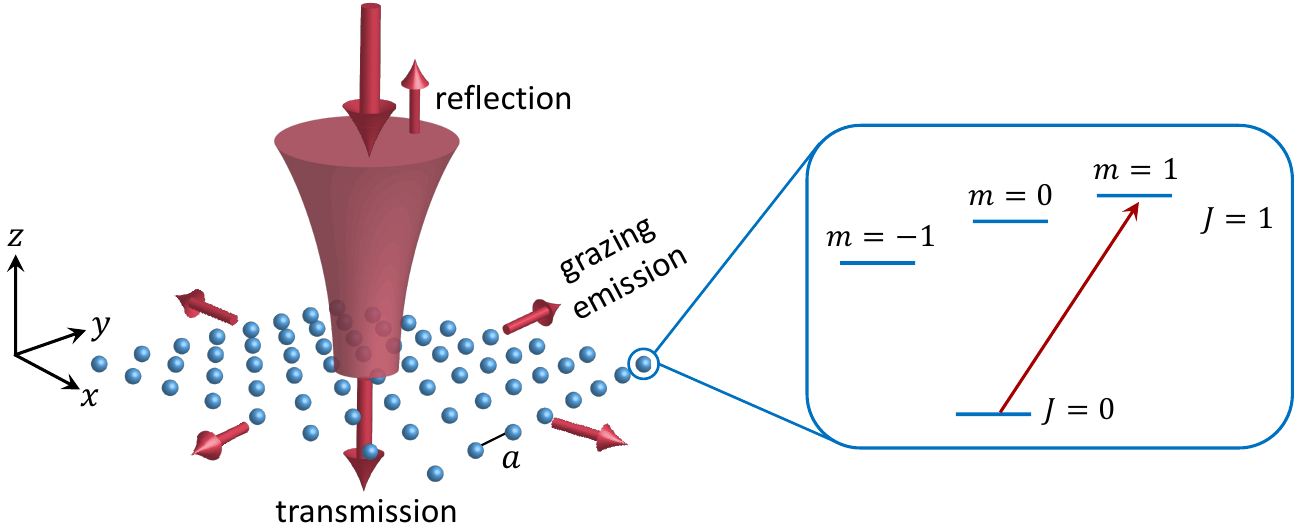}
\centering
\caption{Schematic of light scattering from a two-dimensional atomic array. A finite-waist beam is normally incident on a square array with lattice constant $a$. A magnetic field lifts the degeneracy of the $J=1$ manifold, while the chosen polarization and paraxial excitation isolate an effective two-level transition within the $J=0\rightarrow J=1$ manifold. As $a$ approaches the Rayleigh threshold, the first non-specular diffraction orders enter the light cone and redirect photons into near-grazing modes with $k_z\simeq 0$.
\label{fig:schematic} }
\end{figure}

In this Letter, we show that the Rayleigh anomaly~\cite{Rayleigh1907Dynamical,Hessel1965Wood}, which occurs when a new diffraction order becomes propagating at grazing emission, provides a sharp mechanism for controlling both the linear and nonlinear optical response of a single-layer atomic array. Below the diffraction threshold, the array acts as an efficient atomic mirror~\cite{Bettles2016Enhanced,Shahmoon2017Cooperative,Rui2020Subradiant}. As the first non-specular diffraction orders enter the light cone, the dominant collective modes rapidly broaden, producing an abrupt collapse of the specular reflection and redirecting the scattered light into diffuse off-axis channels. The same threshold provides geometric control over nonlinear two-photon scattering. A $2f$ imaging scheme resolves unnormalized momentum-space coincidences, revealing pronounced opposite-momentum correlations and interference fringes in both reflection and transmission below threshold. As additional diffraction channels open, these structures are strongly reorganized and largely disappear from the collected near-axis modes. Remarkably, in equal-time reflection, the disconnected amplitude exactly cancels the alternating absorption--emission contribution, leaving only the two-excitation response. Its integrated scattering weight collapses by more than three orders of magnitude across the threshold, even after optimizing the incident frequency for single-photon reflection. Meanwhile, transmitted coincidences recover as independent transmission dominates. These findings establish the Rayleigh threshold as a geometric control mechanism for both the strength and joint momentum structure of two-photon scattering in atomic metasurfaces.

\textit{Theoretical framework}---We consider near-resonant photon scattering from a two-dimensional square atomic lattice in the $xy$ plane as shown in Fig.~\ref{fig:schematic}. The array contains $N=L^2$ atoms arranged in an $L\times L$ square lattice with spacing $a$. The dynamics is governed by
$\hat{H}=\hat{H}_{\rm photon}+\hat{H}_{\rm atom}+\hat{H}_{\rm int}$.
The free-space photon Hamiltonian is
$\hat{H}_{\rm photon}=\sum_{\lambda}\int d\boldsymbol{k}\,
\hbar\omega_{\boldsymbol{k}}
\hat{a}^{\dagger}_{\boldsymbol{k},\lambda}
\hat{a}_{\boldsymbol{k},\lambda}$,
where \(\hat{a}_{\boldsymbol{k},\lambda}\) annihilates a photon with wave vector \(\boldsymbol{k}\) and transverse polarization $\lambda$. A magnetic field $\mathbf B=B\mathbf e_z$ isolates the $|J=0,m_J=0\rangle\leftrightarrow|J=1,m_J=+1\rangle$ transition when the Zeeman splitting greatly exceeds the relevant collective shifts, linewidths, and polarization-mixing couplings (see Supplemental Material~\cite{supplementary}). For normally incident circularly polarized light addressing this transition, the atomic dynamics is described by the effective two-level Hamiltonian
$\hat{H}_{\rm atom}=\sum_{j=1}^{N}\hbar\omega_0\hat{\sigma}_j^\dagger\hat{\sigma}_j$~\cite{Bettles2016Enhanced,Olmos2013long},
where $\omega_0$ is the Zeeman-shifted transition frequency, and $\hat{\sigma}_j=|g_j\rangle\langle e_j|$ is the atomic lowering operator. The atom--photon interaction is
$\hat{H}_{\rm int} = \hbar\sum_{j}\sum_{\lambda}\int d\boldsymbol{k}\,
\left[g_{\boldsymbol{k},\lambda}
e^{i\boldsymbol{k}\cdot\boldsymbol{R}_{j}}
\hat{\sigma}^{\dagger}_{j}
\hat{a}_{\boldsymbol{k},\lambda}+{\rm h.c.}
\right]$,
with
$g_{\boldsymbol{k},\lambda} =-i
\sqrt{\omega_{\boldsymbol{k}}/[2\hbar\varepsilon_{0}(2\pi)^{3}]}
 d^{*} \left(
\boldsymbol{e}^{*}_{+}\cdot\boldsymbol{e}_{\boldsymbol{k}\lambda}
\right)$~\cite{supplementary}.
Here \(\boldsymbol{R}_{j}\) is the atomic position, \(\boldsymbol{e}_{\boldsymbol{k}\lambda}\) is the photon polarization vector, and the transition dipole is $\boldsymbol{d}=d\boldsymbol{e}_{+}=d(\boldsymbol{e}_{x}+i\boldsymbol{e}_{y})/\sqrt{2}$.

The linear response is fully determined by the single-photon scattering amplitude $\mathcal{A}^{(1)}(\boldsymbol{p},\lambda';\boldsymbol{k},\lambda)\equiv\langle G,0|
\hat{a}_{\boldsymbol{p},\lambda'}\hat{S}\hat{a}^{\dagger}_{\boldsymbol{k},\lambda}
|0,G\rangle$, where $|0\rangle$ is the photon vacuum and $|G\rangle =\prod_j |g_j\rangle$ is the collective atomic ground state. The incoming and outgoing photon states are connected by the scattering operator $\hat{S}=e^{i\hat{H}_{\rm photon}t_f/\hbar}e^{-i\hat{H}(t_f-t_i)/\hbar}e^{-i\hat{H}_{\rm photon}t_i/\hbar}$
with $t_i\rightarrow -\infty$ and $t_f\rightarrow +\infty$. Using the input-output formalism together with the quantum regression theorem~\cite{Shi2015multiphoton,Caneva2015Quantum,yang2024quantum}, we obtain~\cite{supplementary}
\begin{equation}
\mathcal{A}^{(1)}(\boldsymbol{p},\!\lambda';\!\boldsymbol{k},\!\lambda)
\!=\!
\delta_{\lambda\lambda'}\delta(\boldsymbol{p}-\boldsymbol{k})
-2\pi i\mathcal{T}(\boldsymbol{p},\!\lambda';\!\boldsymbol{k},\!\lambda)
\delta(\omega_{\boldsymbol{p}}-\omega_{\boldsymbol{k}}), \label{eq:amplitude_1p}  \end{equation}
with 
\begin{equation}
\mathcal{T}(\boldsymbol{p},\lambda';\boldsymbol{k},\lambda)
\!=\!\!\sum_{ij}
g^{*}_{\boldsymbol{p},\lambda'}
g_{\boldsymbol{k},\lambda}
e^{-i(\boldsymbol{p}\cdot\boldsymbol{R}_{i}
-\boldsymbol{k}\cdot\boldsymbol{R}_{j})}\mathcal G_{ij}(\omega_{\boldsymbol{k}}),
\end{equation}
and single-excitation Green function
\begin{equation}
\mathcal G_{ij}(\omega)
\equiv
\left\langle G\right|
\hat\sigma_i
\frac{1}{\omega-\hat H_{\rm eff}^{(1)}}
\hat\sigma_j^\dagger
\left|G\right\rangle.\label{eq:Gij2SM}
\end{equation}
Here \(\hat{H}^{(1)}_{\rm eff}\) denotes the single-excitation projection of the effective non-Hermitian atomic Hamiltonian~\cite{AsenjoGarcia2017Selective,Pedersen2024Green}
\begin{equation}
\hat{H}_{\rm eff}
=\sum_{j}
\left(
\omega_{0}-i\frac{\gamma_{0}}{2}
\right)
\hat{\sigma}^{\dagger}_{j}\hat{\sigma}_{j}
+
\sum_{i\neq j}
\left(
J_{ij}-\frac{i}{2}\gamma_{ij}
\right)
\hat{\sigma}^{\dagger}_{i}\hat{\sigma}_{j},    
\end{equation}
which includes single-atom spontaneous decay $\gamma_0$, RDDI strength $J_{ij}$, and cooperative radiative decay $\gamma_{ij}$ induced by vacuum fluctuations~\cite{Agarwal1973RWA,Gruner1996Green,Rusconi2021Photonic}.

The nonlinear response is encoded in the two-photon scattering amplitude $\mathcal{A}^{(2)} =  \left\langle G,0\right|\hat{a}_{\boldsymbol{k}_{4},\lambda_{4}}\hat{a}_{\boldsymbol{k}_{3},\lambda_{3}}\hat{S}\hat{a}^{\dagger}_{\boldsymbol{k}_{2},\lambda_{2}}\hat{a}^{\dagger}_{\boldsymbol{k}_{1},\lambda_{1}}\left|0,G\right\rangle$. It can be decomposed into disconnected and connected parts $\mathcal{A}^{(2)}=\mathcal{A}^{(2)}_{{\rm dis}}+\mathcal{A}^{(2)}_{{\rm con}}$\cite{supplementary,Shi2015multiphoton}, with \begin{equation}
\mathcal{A}^{(2)}_{\rm dis}
=
\left(1+P_{34}\right)
\mathcal{A}^{(1)}(\boldsymbol{k}_3,\lambda_3;\boldsymbol{k}_1,\lambda_1)
\mathcal{A}^{(1)}(\boldsymbol{k}_4,\lambda_4;\boldsymbol{k}_2,\lambda_2),
\end{equation}
where $P_{mn}$ exchanges the labels $m$ and $n$. The disconnected amplitude describes two independent single-photon scattering events, whereas the connected amplitude is the nonfactorizable contribution arising from the intrinsic nonlinear response of the two-level emitters~\cite{supplementary},
\begin{align}
\mathcal{A}^{(2)}_{{\rm con}} &\! =\!-2\pi i\!\!\sum_{j_{1}j_{2}j_{3}j_{4}}\!\!(\!\mathcal{M}_{j_{3}j_{4};j_{1}j_{2}}\!-\!\mathcal{N}_{j_{3}j_{4};j_{1}j_{2}})\delta(\omega_{3}\!+\!\omega_{4}\!-\!\omega_{1}\!-\!\omega_{2})
\nonumber \\
& \times g^{*}_{\boldsymbol{k}_{3},\lambda_{3}}g^{*}_{\boldsymbol{k}_{4},\lambda_{4}}g_{\boldsymbol{k}_{1},\lambda_{1}}g_{\boldsymbol{k}_{2},\lambda_{2}}e^{i(\boldsymbol{k}_{1}\cdot\boldsymbol{R}_{j_{1}}+\boldsymbol{k}_{2}\cdot\boldsymbol{R}_{j_{2}}-\boldsymbol{k}_{3}\cdot\boldsymbol{R}_{j_{3}}-\boldsymbol{k}_{4}\cdot\boldsymbol{R}_{j_{4}})}.
\end{align}
The genuine two-excitation contribution is contained in
\begin{align}
& \mathcal M_{j_{3}j_{4};j_{1}j_{2}}(\omega_{3},\omega_{4};\omega_{1},\omega_{2}) = (1+P_{34})(1+P_{12})\nonumber\\ 
 &\left\langle G\right|\hat{\sigma}_{j_4}
\frac{1}{\omega_4\!-\!\hat{H}^{(1)}_{{\rm eff}}}\hat{\sigma}_{j_3}
\frac{1}{\omega_1\!+\!\omega_2\!-\!\hat{H}^{(2)}_{{\rm eff}}}
\hat{\sigma}^{\dagger}_{j_2}\frac{1}{\omega_1\!-\!\hat{H}^{(1)}_{{\rm eff}}}\hat{\sigma}^{\dagger}_{j_1}\left|G\right\rangle\!,
\end{align}
and \(\hat{H}^{(2)}_{\rm eff}\) is the two-excitation projection of the effective non-Hermitian Hamiltonian. The $\mathcal N$ term originates from the alternating absorption--emission time orderings. Unlike $\mathcal{M}$, it contains no two-excitation propagator and is constrained only by conservation of the total two-photon energy:
\begin{align}
\mathcal N_{j_{3}j_{4};j_{1}j_{2}}&(\omega_{3},\omega_{4};\omega_{1},\omega_{2})  = (1+P_{34})\mathcal P\frac{1}{\omega_3-\omega_1}\nonumber \\ 
 &\times\left[\mathcal{G}_{j_4j_2}(\omega_4)\mathcal{G}_{j_3j_1}(\omega_1)-\mathcal{G}_{j_3j_1}(\omega_3)\mathcal{G}_{j_4j_2}(\omega_2)\right],  
\end{align}
where $\mathcal P$ denotes the Cauchy principal value. Higher-order nonlinear scattering amplitudes can be obtained analogously.

\begin{figure}
\includegraphics[width=8cm]{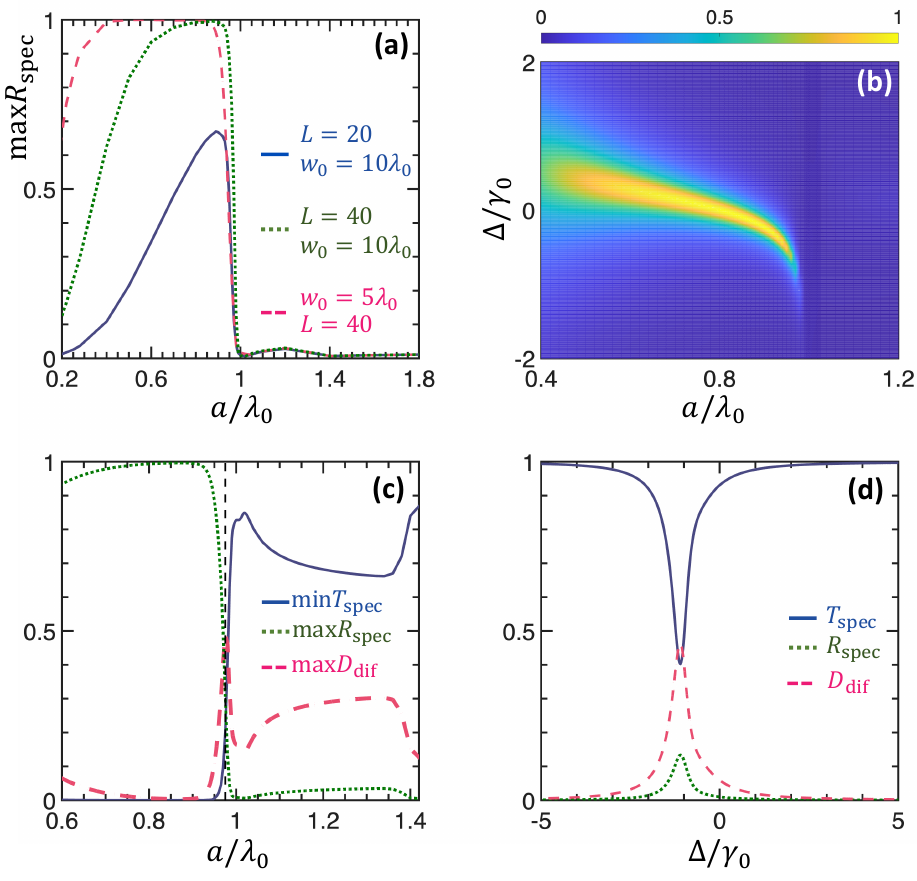}
\centering
\caption{Linear optical response of a finite two-dimensional square atomic array.
(a) Maximum specular reflection rate ${\rm max}R_{\rm spec}$ as a function of the lattice constant $a$ for different array sizes $L$ and Gaussian beam waists $w_0$.
(b) Specular reflection spectrum $R_{\rm spec}$ versus $a$ and detuning $\Delta=\omega_{\rm in}-\omega_0$.
(c) Maximum specular reflection $R_{\rm spec}$, minimum specular transmission $T_{\rm spec}$, and diffuse scattering probability $D_{\rm dif}=1-R_{\rm spec}-T_{\rm spec}$ across the transition. The diffuse scattering exhibits a pronounced peak at $a=0.98\lambda_0$, marked by the vertical dashed line.
(d) Frequency dependence of $T_{\rm spec}$, $R_{\rm spec}$, and $D_{\rm dif}$ at $a=0.98\lambda_0$, showing that resonant scattering is predominantly redistributed into diffuse channels rather than specular reflection.
In panels (b)--(d), the parameters are $L=40$ and $w_0=10\lambda_0$.
 \label{fig:specular_scattering}}
\end{figure}

\textit{Specular scattering and Rayleigh transition}---In contrast to one-dimensional waveguide-QED scattering~\cite{shen2005coherent,shen2005prl,zhou2008controllable}, a two-dimensional atomic array can redistribute photons into a continuum of off-axis modes beyond the specular reflection and transmission channels. Without loss of generality, we consider a left-circularly polarized single-photon pulse $\left|1_{{\rm in}}\right\rangle =\int d\boldsymbol{q}\int^{\infty}_{0}d\omega [A(\boldsymbol{q},\omega)/\sqrt{v_z}]\hat{a}^{\dagger}_{\boldsymbol{q},-k_z,-}\left|0\right\rangle$ incident along the negative $z$ axis with $k_{z}=\sqrt{\omega^{2}/c^{2}-q^{2}}$, transverse wavevector $\boldsymbol{q}=\{k_x,k_y\}$, and effective group velocity $v_{z}=|\partial\omega_{\boldsymbol{k}}/\partial k_{z}|$. We use the single-photon specular reflection and transmission amplitudes to characterize the coherent linear response
\begin{equation}
r_{{\rm spec}} =\left\langle 1_{R}\right|\hat{S}\left|1_{{\rm in}}\right\rangle ,\ t_{{\rm spec}}=\left\langle 1_{T}\right|\hat{S}\left|1_{{\rm in}}\right\rangle,
\end{equation}
Here $|1_T\rangle = |1_{\rm in}\rangle$ and $|1_R\rangle$ has the same spectral distribution as the incident pulse but with reversed $k_z$ and polarization~\cite{supplementary}. Even without nonradiative decay, $R_{\rm spec}=|r_{\rm spec}|^2$ and $T_{\rm spec}=|t_{\rm spec}|^2$ generally sum to less than unity because photons can scatter into diffuse channels.

We focus on quasi-monochromatic paraxial incident pulses, since strong focusing can drive the $\pi$ transition and invalidate the effective two-level description. The pulse spectrum can be factorized as $A(\boldsymbol{q},\omega)\approx\xi(\boldsymbol{q})\chi(\omega)$, with $\int d\boldsymbol{q}|\xi(\boldsymbol{q})|^{2}=\int d\omega|\chi(\omega)|^{2}=1$. In this limit, the specular scattering coefficients reduce to~\cite{supplementary}
\begin{align}
r_{{\rm spec}} & = i\frac{3\pi\gamma_{0}}{2k^{2}_0}\sum_{n}\frac{B_{n}}{\Delta_{n}+i\Gamma_{n}/2},\\
t_{{\rm spec}} & = 1 - i\frac{3\pi\gamma_{0}}{2k^{2}_0}\sum_{n}\frac{B_{n}}{\Delta_{n}+i\Gamma_{n}/2},
\end{align}
which represent a coherent superposition of the scattering spectra of the collective eigenmodes $\hat{H}^{(1)}_{{\rm eff}}\left|n_{R}\right\rangle  =E_{n}\left|n_{R}\right\rangle$. The residue $B_{n}=\sum_{ij}\psi^{*}_{i}U_{in}U_{jn}\psi_{j}$ measures the contribution of the $n$th collective mode to the specular scattering amplitude, determined by its overlap with the incident and detected Gaussian modes $\psi_{i}=(1/2\pi)\int d\boldsymbol{q}\xi(\boldsymbol{q})e^{i\boldsymbol{q}\cdot\boldsymbol{R}_{i}}$, where the $N\times N$ matrix $U$ is composed of the right eigenvectors $|n_R\rangle$. The incident pulse is centered at frequency $\omega_{\rm in}$ with a bandwidth much smaller than $\gamma_0$. Here, $k_0=2\pi/\lambda_0$ is the resonant wavenumber with transition wavelength $\lambda_0=2\pi c/\omega_0$, $\Delta_n=\omega_{\rm in}-{\rm Re}E_n$ denotes the detuning from the collective mode resonance, and $\Gamma_n=-2{\rm Im}E_n$ gives the linewidth of the eigenmode $|n_R\rangle$.

Figure~\ref{fig:specular_scattering}(a) shows that the maximum specular reflection rate ${\rm max}\,R_{\rm spec}$ drops abruptly as the lattice spacing crosses $a/\lambda_0=1$. The incident field is a Gaussian pulse with transverse profile $\xi(\boldsymbol{q})=(w_{0}/\sqrt{2\pi})\exp(-w^{2}_{0}q^{2}/4)$ and beam radius $w_0$, illuminating a square array of side length $L$. Comparing $L=20$ and $L=40$ shows that nearly perfect reflection requires the array to cover the incident beam. The same finite-size effect explains the reduction of ${\rm max}\,R_{\rm spec}$ at smaller $a$. Comparing $w_0=5\lambda_0$ and $10\lambda_0$ further shows that a broader beam sharpens the transition, which becomes increasingly abrupt with increasing beam waist and array size.  Figure~\ref{fig:specular_scattering}(b) shows the cooperative-resonance redshift previously established for infinite lattices~\cite{Shahmoon2017Cooperative}. Crucially, the high-reflectivity resonance terminates near the Rayleigh threshold, leading to an abrupt collapse of the maximum specular reflectance even after optimizing the incident detuning [Fig.~\ref{fig:specular_scattering}(a)]. This behavior is fundamentally distinct from the magic lattice spacing associated with enhanced extinction at fixed detuning reported in Ref.~\cite{Bettles2016Enhanced}. The same threshold-induced collapse is confirmed for a finite triangular lattice near its geometric threshold $a=2\lambda_0/\sqrt{3}$ (see Supplemental Material~\cite{supplementary}), demonstrating that this phenomenon is not specific to square lattices.

Figure~\ref{fig:specular_scattering}(c) shows how the optimal scattering channels evolve across the transition. For each lattice constant, we plot the maximum specular reflection, the minimum specular transmission, and the maximum diffuse scattering probability $D_{\rm dif}=1-R_{\rm spec}-T_{\rm spec}$. Below threshold, the array acts as an efficient mirror with ${\rm max}\,R_{\rm spec}\simeq1$. Close to the transition, however, the specular reflection collapses while the diffuse component develops a pronounced maximum at $a=0.98\lambda_0$, marked by the vertical dashed line. This indicates that the loss of mirror-like reflection is not simply converted into forward transmission, but is instead accompanied by strong redistribution into non-specular radiation channels. Figure~\ref{fig:specular_scattering}(d) shows the corresponding frequency dependence at the lattice spacing $a=0.98\lambda_0$. Near the collective resonance, $T_{\rm spec}$ exhibits a narrow dip while $R_{\rm spec}$ and $D_{\rm dif}$ develop simultaneous Lorentzian peaks.

\begin{figure}
\includegraphics[width=8.5cm]{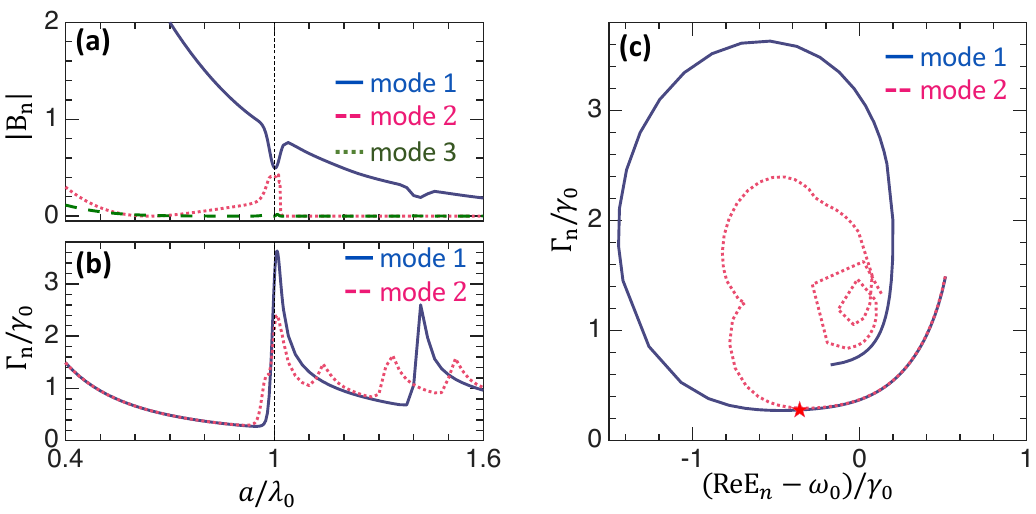}
\centering
\caption{Collective-mode origin of the collapse of specular reflection.
(a) Residues $|B_n|$ of the three collective eigenmodes that couple most strongly to the incident Gaussian beam. (b) Radiative linewidths $\Gamma_n$ of the two leading modes.
(c) Pole trajectories of the two leading modes in the complex spectrum, plotted as $\Gamma_n/\gamma_0$ versus $({\rm Re}E_n-\omega_0)/\gamma_0$. The red star marks the lattice spacing $a=0.93\lambda_0$, where ${\rm max}R_{\rm spec}$ begins to drop. The parameters are $L=40$ and $w_0=10\lambda_0$.
\label{fig:eigenstate_analysis} }
\end{figure}

To identify the microscopic origin of the abrupt loss of specular reflection, we order the collective eigenmodes by the magnitude of their residues $|B_n|$. Figure~\ref{fig:eigenstate_analysis}(a) shows that the response is dominated over most of the parameter range by a single mode with the largest residue $|B_1|$, while other modes couple only weakly to the incident Gaussian beam. Near $a=\lambda_0$, however, $|B_1|$ develops a sharp dip and a second mode acquires finite residue, signaling a redistribution of oscillator strength between two collective resonances. This redistribution is accompanied by strong radiative broadening. As shown in Fig.~\ref{fig:eigenstate_analysis}(b), the linewidths $\Gamma_n$ of the two dominant modes remain relatively small below the transition but increase abruptly at $a/\lambda_0=1$~\cite{Rui2020Subradiant}, suppressing their pole contributions through the denominator $\Delta_n+i\Gamma_n/2$. This diffraction-induced radiative broadening, previously identified for infinite lattices~\cite{Shahmoon2017Cooperative}, is accompanied by a redistribution of radiation into newly opened non-specular channels, thereby suppressing the maximum specular reflectance. The complex resonance poles of these two modes, initially close to each other, separate near the diffraction threshold [Fig.~\ref{fig:eigenstate_analysis}(c)], accompanied by pronounced changes in their resonance frequencies and linewidths. 

This sudden broadening is the Rayleigh anomaly of an effective periodic grating formed by interacting atoms: additional radiative diffraction channels open at grazing emission~\cite{Rayleigh1907Dynamical,Hessel1965Wood}. For an infinitely large array, the eigenstates of $\hat{H}^{(1)}_{\rm eff}$ are Bloch spin waves $\left|\boldsymbol{q}\right\rangle =(1/\sqrt{N})\sum_{j}e^{i\boldsymbol{q}\cdot\boldsymbol{R}_{j}}\hat{\sigma}^{\dagger}_{j}\left|G\right\rangle$. The corresponding radiative linewidth for a square lattice is~\cite{AsenjoGarcia2017Selective}
\begin{equation} \frac{\Gamma_{\boldsymbol{q}}}{\gamma_{0}}=\frac{3\pi}{k^{3}_{0}a^{2}}\sum_{\{\boldsymbol{g}\in|\boldsymbol{q}+\boldsymbol{g}|\leq k_{0}\}}\frac{k^{2}_{0}-|(\boldsymbol{q}+\boldsymbol{g})\cdot\boldsymbol{d}/d|^{2}}{\sqrt{k^{2}_{0}-|\boldsymbol{q}+\boldsymbol{g}|^{2}}}, \end{equation}
where $\boldsymbol{g}=(2\pi/a)(m\boldsymbol{e}_{x}+n\boldsymbol{e}_{y})$ is a reciprocal lattice vector. Below the first diffraction threshold, the $\boldsymbol{q}\simeq
\boldsymbol{0}$ spin waves excited by the normally incident paraxial beam
radiate only through the zeroth-order channel
$\boldsymbol{g}=\boldsymbol{0}$. At $a=\lambda_0$, the first diffraction orders $(m,n)=(\pm1,0),(0,\pm1)$ reach the light cone, $|\boldsymbol{q}+\boldsymbol{g}|\simeq k_0$, where the linewidth denominator becomes singular. These newly opened off-axis channels strongly enhance the radiative linewidth and redirect photons into near-grazing modes with vanishing longitudinal wave number $k^2_z\approx k^{2}_{0}-|\boldsymbol{q}+\boldsymbol{g}|^{2}\simeq0$. The weaker feature near $a=\sqrt{2}\lambda_0$ [see Fig.~\ref{fig:specular_scattering}(c)] comes from the next reciprocal vectors $(m,n)=(\pm1,\pm1)$, which reach the second Rayleigh threshold. For a finite array driven by a finite-waist beam, the singularities are rounded into the sharp linewidth peaks in Fig.~\ref{fig:eigenstate_analysis}(b), causing the collapse of specular reflection.

\begin{figure}
\includegraphics[width=8.5cm]{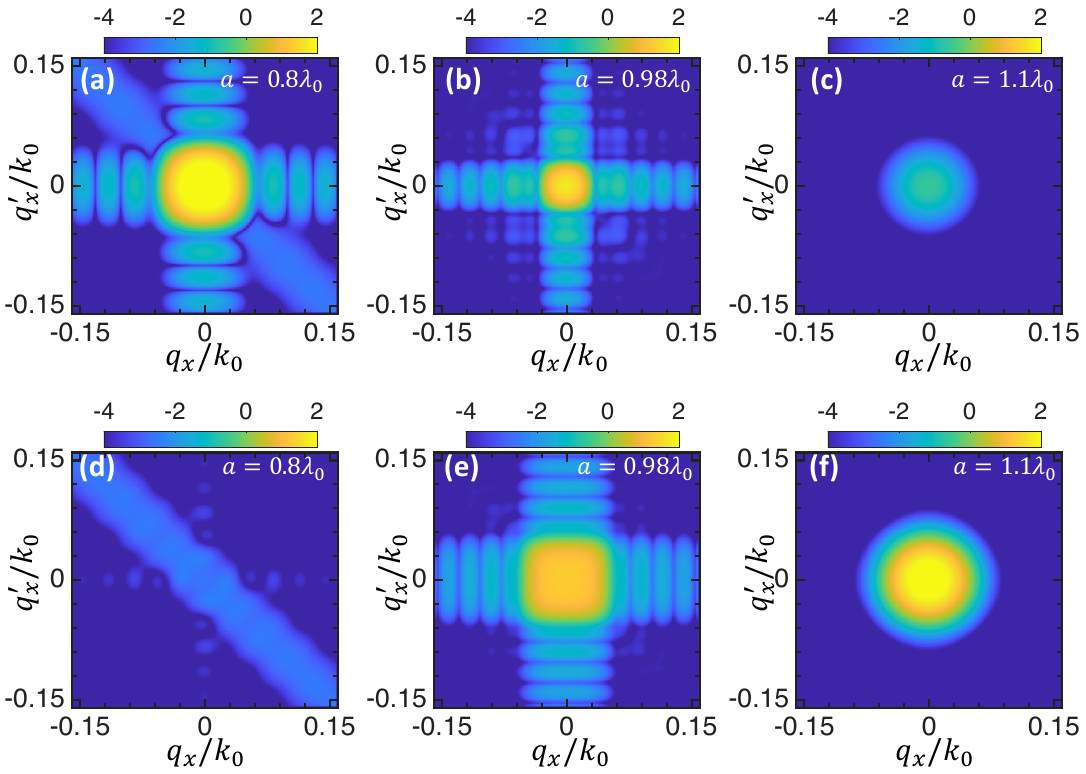}
\centering
\caption{Momentum-space two-photon correlations across the Rayleigh threshold. Equal-time unnormalized coincidence signals $\log_{10}G_{\alpha\alpha}^{(2)}(q_x,0;q_x',0)$ for reflected photons ($\alpha=R$, top row) and transmitted photons ($\alpha=T$, bottom row), measured in a $2f$ configuration. The three columns correspond to $a=0.8\lambda_0$, $0.98\lambda_0$, and $1.1\lambda_0$, respectively. Below threshold, pronounced opposite-momentum correlations and interference fringes are observed, which are reorganized near the threshold and largely disappear above it. At each lattice spacing, the detuning is chosen to maximize the single-photon specular reflectance, as in Fig.~\ref{fig:specular_scattering}(a). The parameters are $L=40$ and $w_0=10\lambda_0$, with a collection aperture $|\boldsymbol q|<0.16k_0$. \label{fig:correlation}}
\end{figure}

\textit{Momentum-space photon correlations.}---The Rayleigh threshold controls both the strength and joint momentum
distribution of scattered photon pairs.
A $2f$ lens system maps transverse momentum onto positions in the
detection plane, $\boldsymbol q=k_0\boldsymbol\rho/f$, enabling
momentum-resolved measurements of single-photon scattering and
two-photon correlations~\cite{supplementary}. We restrict detection to a low-NA cone ($|\boldsymbol q/k_0|<1/2\pi\approx 0.16$), justifying the paraxial and scalar-field approximations. Unlike the collective dipole blockade characterized by time-dependent, angle-integrated photon correlations in Ref.~\cite{Williamson2020Superatom}, we investigate the equal-time, unnormalized, momentum-resolved two-photon coincidence signals in reflection and transmission, revealing their restructuring across the Rayleigh threshold
\begin{equation}
G^{(2)}_{\alpha\alpha}(\boldsymbol\rho,\boldsymbol\rho')
\!=\!
\left|
\langle 0|
\hat E^{(+)}_\alpha(\boldsymbol\rho')
\hat E^{(+)}_\alpha(\boldsymbol\rho)
|\Psi^{(2)}_{\rm out}\rangle
\right|^2
\!\equiv\!
\left|
\mathcal A^{(2)}_{\alpha\alpha}
(\boldsymbol\rho,\boldsymbol\rho')
\right|^2,
\end{equation}
where $\alpha=R,T$ and $\hat E^{(+)}_\alpha$ denotes the positive-frequency field operator in the detection plane.
Although the near-grazing diffraction orders lie outside the collection aperture, their opening modifies the two-photon
scattering into the detected near-axis modes.

The two-photon amplitude separates into three contributions $\mathcal A^{(2)}_{\alpha\alpha} =\mathcal A^{(2)}_{\alpha\alpha,\mathrm{dis}} +\mathcal A^{(2)}_{\alpha\alpha,M} +\mathcal A^{(2)}_{\alpha\alpha,N}$. The disconnected term describes independent single-photon scattering, whereas the connected $\mathcal M$ and $\mathcal N$ terms arise from two-excitation propagation and alternating absorption--emission processes (the principal-value contribution), respectively. For equal-time, frequency-integrated detection within the narrowband Markov approximation, the $\mathcal N$ term exactly cancels the disconnected contribution in reflection, yielding $\mathcal A^{(2)}_{RR}=\mathcal A^{(2)}_{RR,M}$~\cite{supplementary}. To quantify the absolute two-photon scattering weight, we define
\begin{equation}
W_{\alpha\alpha}
=\int_{\rm NA} d^2\boldsymbol q
\int_{\rm NA} d^2\boldsymbol q'\,
\left|
\mathcal A^{(2)}_{\alpha\alpha}
(\boldsymbol q,\boldsymbol q')
\right|^2.
\label{eq:W_RT}
\end{equation}
Thus, $W_{RR}$ measures the integrated two-excitation contribution without normalization by single-photon intensities.

Figure~\ref{fig:correlation} reveals a striking restructuring of the joint momentum distribution across the Rayleigh threshold. Below threshold $a=0.8\lambda_0$, both reflected and transmitted photons exhibit pronounced opposite-momentum correlations along $q_{x}+q'_{x}\simeq0$, accompanied by finite-array interference fringes~\cite{supplementary}. The opposite-momentum structure arises from coherent scattering by collectively excited atomic pairs, whose center-of-mass coordinate is conjugate to the total transverse photon momentum. Near threshold $a=0.98\lambda_0$, these correlations are reorganized into intricate interference patterns. Above threshold $a=1.1\lambda_0$, the reflected coincidence signal is strongly suppressed, while transmitted coincidences are dominated by a smooth, nearly separable background.

\begin{figure}
\includegraphics[width=8.5cm]{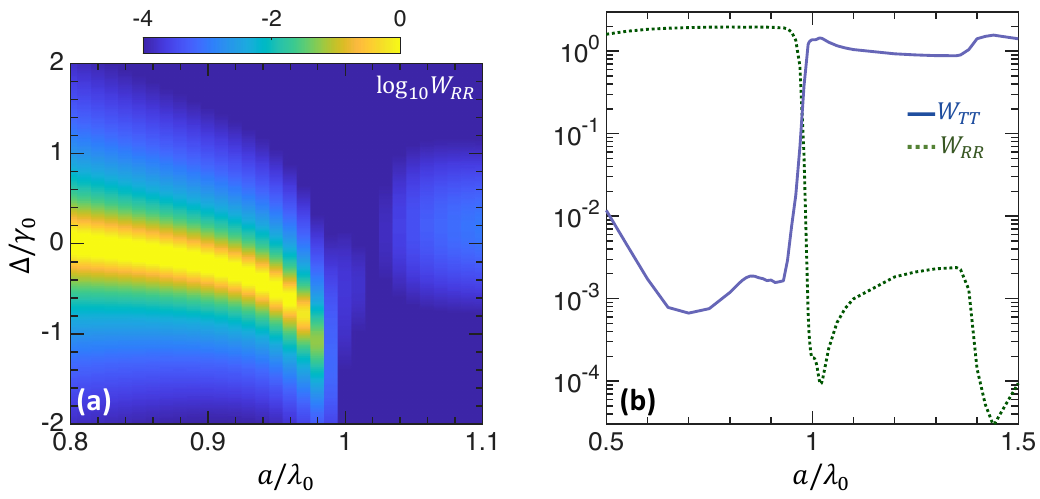}
\centering
\caption{Collapse of two-photon scattering across the Rayleigh threshold. (a) Integrated reflected two-photon weight $\log_{10}W_{RR}$ as a function of lattice spacing $a$ and detuning $\Delta$. (b) Integrated reflected ($W_{RR}$) and transmitted ($W_{TT}$) two-photon weights, with the detuning optimized to maximize single-photon specular reflection at each $a$. The reflected weight drops by more than three orders of magnitude near $a=\lambda_0$, while transmission recovers. Parameters: $L=40$, $w_0=10\lambda_0$, and $|\boldsymbol q|<0.16k_0$.\label{fig:integrateW}}
\end{figure}

Crucially, this restructuring is accompanied by a suppression of the absolute two-excitation scattering weight $W_{RR}$ by more than three orders of magnitude. Figure~\ref{fig:integrateW}(a) maps $W_{RR}$, which equals the integrated $\mathcal M$-term weight, as a function of lattice spacing and detuning. Its high-weight region collapses sharply near $a=\lambda_0$.
This suppression persists even when the detuning is optimized to maximize single-photon specular reflection at each lattice
spacing [Fig.~5(b)], whereas $W_{TT}$ recovers as direct transmission becomes dominant. The opening of additional diffraction channels thus redirects collective radiation and reshapes the two-excitation response, suppressing correlated pair scattering into the collected modes. Together, Figs.~\ref{fig:correlation} and~\ref{fig:integrateW} demonstrate geometric control of both the momentum structure and absolute weight of two-photon scattering, beyond effects arising solely from intensity normalization.

\textit{Implementation}.—A possible realization employs a two-dimensional array of $^{88}$Sr atoms on the $^1S_0\rightarrow{}^3P_1$ transition ($\lambda_0=689$ nm, $\gamma_0/2\pi\simeq7.5$ kHz), building on demonstrated strontium optical-tweezer and lattice arrays~\cite{Norcia2018SrTweezers,Tao2024SrLatticeArrays}. A magnetic field $\mathbf B=B\mathbf e_z$ perpendicular to the array lifts the degeneracy of the $J=1$ manifold, allowing circularly polarized light to selectively address the $m_J=+1$ excited state. Magnetic control of the $^3P_1$ Zeeman sublevels has been demonstrated experimentally~\cite{Winchester2017SrZeeman}. At $B=1$ G, the Zeeman splitting between adjacent levels is $\delta_Z/2\pi\simeq2.1$ MHz ($\delta_Z\simeq280\gamma_0$), far exceeding the linewidths of the dominant collective modes. This large splitting suppresses off-resonant polarization mixing, supporting the effective two-level description as further examined in the Supplemental Material.

\textit{Conclusion}---We have demonstrated that the Rayleigh diffraction threshold provides geometric control over both linear scattering and nonlinear two-photon correlations in atomic arrays. The opening of additional radiative channels suppresses specular reflection and strongly reorganizes the joint photon momentum distribution, causing pronounced opposite-momentum correlations to disappear above threshold. Remarkably, the disconnected reflected amplitude exactly cancels the alternating absorption--emission contribution, leaving only the genuine two-excitation response. Its integrated scattering weight collapses by more than three orders of magnitude, even after optimizing the incident detuning for single-photon reflection. These effects, accessible through momentum-resolved coincidence measurements in a $2f$ configuration, establish Rayleigh diffraction thresholds as a geometric means of controlling nonlinear photon scattering in atomically thin quantum metasurfaces. This approach may also be extended to cavity-assisted generation of correlated light~\cite{Liu2023Realization,Yan2023Superradiant,Seubert2025tweezer,Zhang2024Cavity}.

\section*{Acknowledgment}
This work is funded by the Quantum Science and Technology-National Science and Technology Major Project (No.~2023ZD0300700), the National Natural Science Foundation of China (NSFC) Grant No. 12275048, and the Fundamental Research Funds for the Central Universities (No. 2412025QD010).

\vspace{-0.5cm}
\bibliography{main}
\end{document}


\title{Supplementary Material for "Rayleigh Diffraction Threshold Controls Nonlinear Photon Correlations in Atomic Arrays"}

\author{Zai-Cheng Xiao}
\author{Li-Ping Yang}
\email{lipingyang87@gmail.com}
\affiliation{Center for Quantum Sciences and School of Physics, Northeast Normal University, Changchun 130024, China}

\maketitle

\section{Hamiltonian and effective atomic dynamics}
\label{sec:SMHamiltonian}

We consider $N$ identical effective two-level emitters located at $\boldsymbol R_j=(x_j,y_j,0)$, with $\hat\sigma_j=|g_j\rangle\langle e_j|$.  The transition frequency $\omega_0$ includes the isolated-atom Lamb shift.  The microscopic Hamiltonian is $\hat H=\hat H_{\rm atom}+\hat H_{\rm photon} +\hat H_{\rm int}$, with
\begin{align}
\hat H_{\rm atom}
 &=\hbar\omega_0\sum_{j=1}^N\hat\sigma_j^\dagger\hat\sigma_j,
 \label{eq:HatomSM}\\
\hat H_{\rm photon}
 &=\hbar\sum_\lambda\int d\boldsymbol k\,
 \omega_{\boldsymbol k}\hat a_{\boldsymbol k,\lambda}^\dagger
 \hat a_{\boldsymbol k,\lambda},\qquad \omega_{\boldsymbol k}=c|\boldsymbol k|,
 \label{eq:HphSM}\\
\hat H_{\rm int}
 &=\hbar\sum_{j,\lambda}\int d\boldsymbol k\,
 \left[g_{\boldsymbol k,\lambda}
 e^{i\boldsymbol k\cdot\boldsymbol R_j}
 \hat\sigma_j^\dagger\hat a_{\boldsymbol k,\lambda}
 +\mathrm{h.c.}\right],
 \label{eq:HintSM}\\
g_{\boldsymbol k,\lambda}
 &=-i\sqrt{\frac{\omega_{\boldsymbol k}}
 {2\hbar\varepsilon_0(2\pi)^3}}\,
 \boldsymbol e_{\boldsymbol k\lambda}^*\cdot\boldsymbol d,
 \qquad \boldsymbol d=d\boldsymbol e_+,
 \quad\boldsymbol e_+=\frac{\boldsymbol e_x+i\boldsymbol e_y}{\sqrt2}.
 \label{eq:gklambda}
\end{align}
Here $\lambda$ denotes either transverse polarization for each three-dimensional wavevector $\boldsymbol k$.  The interaction in Eq.~\eqref{eq:HintSM} is the excitation-conserving rotating-wave approximation (RWA) to the electric-dipole coupling, with $\hat{\boldsymbol d}_j=\boldsymbol d^*\hat\sigma_j^\dagger +\boldsymbol d\hat\sigma_j$.

For wave vector $\boldsymbol{k}=k(\sin\theta\cos\phi,\sin\theta\sin\phi,\cos\theta)$, the unit vectors of the two transverse polarizations in the lab frame are given by
\begin{align}
\boldsymbol{e}_{\boldsymbol{k}\theta} & =\cos\theta\cos\phi\boldsymbol{e}_{x}+\cos\theta\sin\phi\boldsymbol{e}_{y}-\sin\theta\boldsymbol{e}_{z}\\
\boldsymbol{e}_{\boldsymbol{k}\phi} & =-\sin\phi\boldsymbol{e}_{x}+\cos\phi\boldsymbol{e}_{y}.
\end{align}
The polarization vector for circularly polarized modes are 
\begin{equation}
\boldsymbol{e}_{\boldsymbol{k},\pm}=\frac{1}{\sqrt{2}}\left(\boldsymbol{e}_{\boldsymbol{k}\theta}\pm i\boldsymbol{e}_{\boldsymbol{k}\phi}\right).    
\end{equation}

\subsection{Born--Markov master equation and the rotating-wave approximation}
\label{sec:3DBM}

Eliminating the three-dimensional vacuum electromagnetic field from the electric-dipole interaction, followed by the Born--Markov and
optical secular approximations, gives the standard collective master equation~\cite{Lehmberg1970Radiation,Agarwal1973RWA,Dung2002Resonant,Gruner1996Green}:
\begin{align}
\frac{d\hat\rho}{dt}
={}&-\frac{i}{\hbar}[\hat H_{\rm atom}+\hat H_{\rm RDDI},\hat\rho]
+\sum_{i,j}\gamma_{ij}\left(\hat\sigma_j\hat\rho\hat\sigma_i^\dagger
-\frac12\{\hat\sigma_i^\dagger\hat\sigma_j,\hat\rho\}\right),
\label{eq:MasterSM}
\end{align}
with $\hat H_{\rm RDDI}
=\hbar\sum_{i\ne j}J_{ij}\hat\sigma_i^\dagger\hat\sigma_j$. The coherent resonant dipole--dipole interaction (RDDI) and the collective radiative decay are expressed through the outgoing-wave free-space dyadic Green tensor as
\begin{align}
J_{ij}
&=-\frac{\omega_0^2}{\hbar\varepsilon_0c^2}
\boldsymbol d^*\cdot\operatorname{Re}
\overleftrightarrow G(\boldsymbol R_i,\boldsymbol R_j;\omega_0)
\cdot\boldsymbol d,\quad i\ne j,
\label{eq:JGreenSM}\\
\gamma_{ij}
&=\frac{2\omega_0^2}{\hbar\varepsilon_0c^2}
\boldsymbol d^*\cdot\operatorname{Im}
\overleftrightarrow G(\boldsymbol R_i,\boldsymbol R_j;\omega_0)
\cdot\boldsymbol d,
\label{eq:gammaGreenSM}
\end{align}
where the on-site Lamb shift is absorbed into $\omega_0$ and $\gamma_{ii}=\gamma_0=\omega_0^3|d|^2/(3\pi\varepsilon_0\hbar c^3)$.
To obtain the complete resonant dipole--dipole interaction (RDDI), the counter-rotating terms in the microscopic interaction
Hamiltonian $\hat H_{\rm int}$ must, in principle, be retained. These terms contribute to the coherent coupling $J_{ij}$ through
virtual photon exchange and also generate excitation-nonconserving terms in the master equation, such as $\hat\sigma_i\hat\sigma_j$ and
$\hat\sigma_i^\dagger\hat\sigma_j^\dagger$. The latter rapidly oscillating terms are subsequently discarded under the secular approximation, while their virtual contributions to the excitation-conserving RDDI are retained \cite{Agarwal1973RWA,Dung2002Resonant}.

It is also useful to clarify the frequently employed microscopic RWA. With the vacuum spectral density
\begin{equation}
\mathcal J_{ij}(\omega)=\frac{\nu^2}{\pi\hbar\varepsilon_0c^2}
\boldsymbol d^*\cdot\operatorname{Im}
\overleftrightarrow G(\boldsymbol R_i,\boldsymbol R_j;\omega)
\cdot\boldsymbol d,\qquad \nu>0,
\label{eq:JrhoGreenSM}
\end{equation}
Born--Markov elimination of the RWA interaction alone gives
\begin{equation}
\gamma_{ij}=2\pi\mathcal J_{ij}(\omega_0),\qquad
J_{ij}^{\rm RWA}=\mathcal P\!\int_0^\infty\!d\omega\,
\frac{\mathcal J_{ij}(\nu)}{\omega_0-\omega},
\label{eq:JpositiveGreenSM}
\end{equation}
so RWA does produce coherent RDDI, but its positive-frequency principal-value integral is not, in general, exactly equal to Eq.~\eqref{eq:JGreenSM}.  For identical transitions, the omitted counter-rotating virtual processes supply the complementary dispersive contribution, giving 
\begin{equation}
J_{ij}=\mathcal P\int_0^\infty d\omega\,\mathcal J_{ij}(\nu)
\left(\frac{1}{\omega_0-\omega}-\frac{1}{\omega_0+\omega}\right),
\qquad i\ne j,
\label{eq:JfullPVSM}
\end{equation}
which yields Eq.~\eqref{eq:JGreenSM} by the Kramers--Kronig relation. Alternatively, the conventional optical-frequency prescription extends the RWA principal-value integral to negative frequencies, with the appropriate Green-tensor frequency symmetry~\cite{Pedersen2024Green}. This is an approximation, not an exact identity of the bare RWA Hamiltonian.  It is appropriate when off-resonant corrections are negligible for the near-resonant optical observables (in particular, $\omega_0\gg\gamma_0,|J_{ij}|$ and interatomic separations are not deeply subwavelength, as near the threshold $a\sim\lambda_0$). The coefficients used here are those of Eqs.~\eqref{eq:JGreenSM} and \eqref{eq:gammaGreenSM}, consistently obtainable from the full dipole coupling followed by secularization.  The RWA used below for near-resonant photon scattering does not discard the virtual-process contributions already included in $J_{ij}$.

\subsection{Quantum regression and the non-Hermitian propagator}
\label{sec:QRTSM}

For the scattering calculation we write the same master equation as
\begin{align}
\dot{\hat\rho}&=-i(\hat H_{\rm eff}\hat\rho
-\hat\rho\hat H_{\rm eff}^\dagger)
+\mathcal L_{\rm jump}[\hat\rho],
\label{eq:masterUnravelSM}\\
\hat H_{\rm eff}&=\omega_0\sum_j\hat\sigma_j^\dagger\hat\sigma_j
+\sum_{i\ne j}J_{ij}\hat\sigma_i^\dagger\hat\sigma_j
-\frac{i}{2}\sum_{i,j}\gamma_{ij}
\hat\sigma_i^\dagger\hat\sigma_j,
\label{eq:HeffSM}\\
\mathcal L_{\rm jump}[\hat\rho]&=
\sum_{i,j}\gamma_{ij}\hat\sigma_j\hat\rho\hat\sigma_i^\dagger.
\label{eq:jumpSM}
\end{align}
Here $\hat H_{\rm eff}$ is in frequency units.  The quantum regression theorem expresses the vacuum two-time correlator as
\begin{equation}
\langle G|\hat\sigma_i(\tau)\hat\sigma_j^\dagger(0)|G\rangle
=\operatorname{Tr}\!\left[\hat\sigma_i e^{\mathcal L\tau}
(\hat\sigma_j^\dagger|G\rangle\langle G|)\right],
\qquad \tau\ge0,
\label{eq:QRTdefSM}
\end{equation}
with $|G\rangle=\prod_j|g_j\rangle$.
Since $\mathcal L_{\rm jump}[|j\rangle\langle G|]=0$ for $|j\rangle=\hat\sigma_j^\dagger|G\rangle$, the quantum regression theorem reduces to
\begin{equation}
\langle\hat\sigma_i(\tau)\hat\sigma_j^\dagger(0)\rangle_G
=\langle i|e^{-i\hat H_{\rm eff}^{(1)}\tau}|j\rangle,
\label{eq:QRTpropSM}
\end{equation}
which supplies the resolvent $[\omega-\hat H_{\rm eff}^{(1)}]^{-1}$ in the single-photon free-space scattering amplitude below.  The same causal
no-jump propagation applies between field insertions in vacuum multiphoton scattering correlators~\cite{yang2024quantum,
Shi2015multiphoton,Caneva2015Quantum}. It is not a statement that jumps vanish in general atomic density-matrix dynamics.

\section{Input--output relation and photon scattering amplitudes}
\label{sec:SMinputoutput}

\subsection{Three-dimensional free-space input--output relation}

The Heisenberg equation for every free-space mode is
\begin{equation}
\frac{d}{dt}\hat a_{\boldsymbol k,\lambda}(t)
=-i\omega_{\boldsymbol k}\hat a_{\boldsymbol k,\lambda}(t)
-i\sum_jg^*_{\boldsymbol k,\lambda}\hat\sigma_j(t)
 e^{-i\boldsymbol k\cdot\boldsymbol R_j}.
\label{eq:adotSM}
\end{equation}
Solving Eq.~\eqref{eq:adotSM} from the initial time $t_i$ to an intermediate time $t$, and from $t$ to the final time $t_f$, respectively, yields
\begin{align}
\hat a_{\boldsymbol k,\lambda}(t)e^{i\omega_{\boldsymbol k}t}
&=\hat a_{\boldsymbol k,\lambda}(t_i)e^{i\omega_{\boldsymbol k}t_i}
-i\sum_jg^*_{\boldsymbol k,\lambda}e^{-i\boldsymbol k\cdot\boldsymbol R_j}
\int_{t_i}^{t}ds\,\hat\sigma_j(s)e^{i\omega_{\boldsymbol k}s},\\
\hat a_{\boldsymbol k,\lambda}(t_f)e^{i\omega_{\boldsymbol k}t_f}
&=\hat a_{\boldsymbol k,\lambda}(t)e^{i\omega_{\boldsymbol k}t}
-i\sum_jg^*_{\boldsymbol k,\lambda}e^{-i\boldsymbol k\cdot\boldsymbol R_j}
\int_{t}^{t_f}ds\,\hat\sigma_j(s)e^{i\omega_{\boldsymbol k}s}
\end{align}
We define the input, output, and source operators as
\begin{equation}
\hat{a}^{\rm in}_{\boldsymbol{k},\lambda} = \hat a_{\boldsymbol k,\lambda}(t_i)e^{i\omega_{\boldsymbol k}t_i},\qquad  \hat{a}^{\rm out}_{\boldsymbol{k},\lambda} = \hat a_{\boldsymbol k,\lambda}(t_f)e^{i\omega_{\boldsymbol k}t_f}, \qquad \hat{\mathcal S}_{\boldsymbol k,\lambda} = -i\sum_jg^*_{\boldsymbol k,\lambda}e^{-i\boldsymbol k\cdot\boldsymbol R_j}
\int_{t_i}^{t_f}dt\,\hat\sigma_j(t)e^{i\omega_{\boldsymbol k}t},\label{eq:IOS_operators}
\end{equation}
with commutation relations $[\hat{a}_{\boldsymbol{k},\lambda}^{\rm in},\hat{a}_{\boldsymbol{k}',\lambda'}^{\rm in\dagger}]=[\hat{a}_{\boldsymbol{k},\lambda}^{\rm out},\hat{a}_{\boldsymbol{k}',\lambda'}^{\rm out\dagger}]=\delta_{\lambda\lambda'}\delta(\boldsymbol{k}-\boldsymbol{k})$. Taking the scattering limit $t_i\to-\infty$ and
$t_f\to+\infty$, we obtain the input--output relation
\begin{equation}
\hat{a}^{\rm out}_{\boldsymbol{k},\lambda}=\hat{a}^{\rm in}_{\boldsymbol{k},\lambda}+\hat{\mathcal S}_{\boldsymbol k,\lambda}.
\label{eq:IO_SM}
\end{equation}
This is an exact operator identity within the rotating-wave Hamiltonian, without invoking the Markov approximation.

For the subsequent evaluation of photon-cattering amplitudes, it is useful to express the input and output operators in terms of the field and atomic operators at an intermediate time:
\begin{align}
\hat{a}^{\rm in}_{\boldsymbol{k},\lambda} & = \hat{a}_{\boldsymbol k,\lambda}(t)e^{i\omega_{\boldsymbol k}t} +i\sum_jg^*_{\boldsymbol k,\lambda}e^{-i\boldsymbol k\cdot\boldsymbol R_j}
\int_{t_i}^{t}ds\hat\sigma_j(s)e^{i\omega_{\boldsymbol k}s},\\
\hat{a}^{\rm out}_{\boldsymbol{k},\lambda} & = \hat{a}_{\boldsymbol k,\lambda}(t)e^{i\omega_{\boldsymbol k}t} -i\sum_jg^*_{\boldsymbol k,\lambda}e^{-i\boldsymbol k\cdot\boldsymbol R_j}
\int_{t}^{t_f}ds\hat\sigma_j(s)e^{i\omega_{\boldsymbol k}s}.
\end{align}
Since the field and atomic operators commute at equal times, $[\hat a_{\boldsymbol k,\lambda}(t),\hat O(t)]=0$ for any atomic operator $\hat O(t)$, these identities imply the following unequal-time commutation relations:
\begin{align}
\left[\hat{a}^{\rm in}_{\boldsymbol{k},\lambda},\hat{O}(t)\right] = i\sum_jg^*_{\boldsymbol k,\lambda}e^{-i\boldsymbol k\cdot\boldsymbol R_j}
\int_{t_i}^{t}dse^{i\omega_{\boldsymbol k}s}\left[\hat\sigma_j(s),\hat{O}(t)\right],\label{eq:CR_ain} \\
\left[\hat{a}^{\rm out}_{\boldsymbol{k},\lambda},\hat{O}(t)\right] = -i\sum_jg^*_{\boldsymbol k,\lambda}e^{-i\boldsymbol k\cdot\boldsymbol R_j}
\int_{t}^{t_f}ds e^{i\omega_{\boldsymbol k}s}\left[\hat\sigma_j(s),\hat{O}(t)\right].\label{eq:CR_aout}
\end{align}
These relations allow the field-operator contributions to the scattering amplitudes to be expressed in terms of atomic correlation functions, while explicitly preserving their causal time ordering.

\subsection{Single-photon scattering amplitude}
The single-photon scattering amplitude is defined as
\begin{align}
\mathcal A^{(1)}(\boldsymbol{p},\lambda';\boldsymbol{k},\lambda)
& = \langle G,0|\hat a_{\boldsymbol{p},\lambda'}\hat S
\hat a_{\boldsymbol{k},\lambda}^\dagger|0,G\rangle = e^{i(\omega_{\boldsymbol{p}} t_{f}-\omega_{\boldsymbol{k}}t_{i})}\left\langle G,0\right|\hat{a}_{\boldsymbol{p},\lambda'}(t_{f})\hat{a}^{\dagger}_{\boldsymbol{k},\lambda}(t_{i})\left|0,G\right\rangle,
\label{eq:A1defSM}
\end{align}
where
\begin{equation}
\hat S=e^{i\hat H_{\rm photon}t_f/\hbar}
e^{-i\hat H(t_f-t_i)/\hbar}
e^{-i\hat H_{\rm photon}t_i/\hbar}
\end{equation}
is the scattering operator, $|0\rangle$ is the vacuum state of photons, and $|G\rangle=\prod_j|g_j\rangle$.  

Using the input- and output-field operators defined in Eq.~\eqref{eq:IOS_operators}, together with the
input--output relation in Eq.~\eqref{eq:IO_SM},
we obtain
\begin{align}
\mathcal A^{(1)}(\boldsymbol{p},\lambda';\boldsymbol{k},\lambda) & = \left\langle G,0\right|\hat{a}_{\boldsymbol{p},\lambda'}^{\rm out}\hat{a}^{\rm in\dagger}_{\boldsymbol{k},\lambda}\left|0,G\right\rangle = \left\langle G,0\right|\left(\hat{a}_{\boldsymbol{p},\lambda'}^{\rm in}+\hat{\mathcal S}_{\boldsymbol{p},\lambda'}\right)\hat{a}^{\rm in\dagger}_{\boldsymbol{k},\lambda}\left|0,G\right\rangle \\
& = \delta_{\lambda\lambda'}\delta(\boldsymbol{p}-\boldsymbol{k}) +\left\langle G,0\right|\left(\hat{a}^{\rm in\dagger}_{\boldsymbol{k},\lambda}\hat{\mathcal S}_{\boldsymbol{p},\lambda'}-[\hat{a}^{\rm in\dagger}_{\boldsymbol{k},\lambda},\hat{\mathcal S}_{\boldsymbol{p},\lambda'}]\right)\left|0,G\right\rangle.
\end{align}
For the excitation-conserving total Hamiltonian, the collective ground state and photon vacuum form a stationary state, satisfying
\begin{equation}
 \hat{a}^{\rm in}_{\boldsymbol{k},\lambda}|0,G\rangle = \hat{a}^{\rm out}_{\boldsymbol{k},\lambda}|0,G\rangle =\hat{\sigma}_j(t)|0,G\rangle =0.   
\end{equation}
Using the Hermitian conjugate of Eq.~\eqref{eq:CR_ain}, we find
\begin{equation}
\mathcal A^{(1)}(\boldsymbol{p},\lambda';\boldsymbol{k},\lambda)= \delta_{\lambda\lambda'}\delta(\boldsymbol{p}-\boldsymbol{k}) -\sum_{ij}g^{*}_{\boldsymbol{p},\lambda'}g_{\boldsymbol{k},\lambda}e^{-i(\boldsymbol{p}\cdot\boldsymbol{R}_{i}-\boldsymbol{k}\cdot\boldsymbol{R}_{j})}\int^{t_{f}}_{t_{i}}dt_{2}\int^{t_{2}}_{t_{i}}dt_{1}\left\langle G,0\right|\hat{\sigma}_{i}(t_{2})\hat{\sigma}^{\dagger}_{j}(t_{1})\left|0,G\right\rangle e^{i(\omega_{\boldsymbol{p}}t_{2}-\omega_{\boldsymbol{k}}t_{1})}.    
\end{equation}
where we used the vacuum expectation value of the commutator
\begin{equation}
\langle G,0|[\hat{\sigma}^{\dagger}_j(t_1),\hat{\sigma}_i(t_2)]|0,G\rangle = -\langle G,0|\hat{\sigma}_i(t_2)\hat{\sigma}^{\dagger}_j(t_1)|0,G\rangle.    
\end{equation}

Since the collective ground state is stationary, the two-time atomic correlation function depends only on the time difference ($\tau=t_2-t_1$). Within the Born--Markov approximation, the quantum regression theorem, the time-ordered atomic correlation function can be evaluated via the quantum regression theorem~\cite{yang2024quantum,Shi2015multiphoton,Caneva2015Quantum}
\begin{equation}
\langle G,0|\hat{\sigma}_i(t_2)\hat{\sigma}^{\dagger}_j(t_1)|0,G\rangle =\langle G|\hat\sigma_i e^{-i\hat H_{\mathrm{eff}}^{(1)}\tau}
\hat\sigma_j^\dagger|G\rangle.
\end{equation}
where $\hat H_{\mathrm{eff}}^{(1)}$ denotes the single-excitation projection of the effective
non-Hermitian Hamiltonian defined in Eq.~\eqref{eq:HeffSM}.

Taking the scattering limit $t_i\to-\infty$ and $t_f\to+\infty$, we obtain
\begin{equation}
\mathcal A^{(1)}(\boldsymbol{p},\lambda';
\boldsymbol{k},\lambda)
=
\delta_{\lambda\lambda'}\delta(\boldsymbol{p}-\boldsymbol{k})
-2\pi i\,
\mathcal T(\boldsymbol{p},\lambda';
\boldsymbol{k},\lambda)
\delta(\omega_{\boldsymbol{p}}-\omega_{\boldsymbol{k}}),
\label{eq:A1SM}
\end{equation}
where
\begin{align}
\mathcal T(\boldsymbol{p},\lambda';
\boldsymbol{k},\lambda)
&=
\sum_{ij}
g^*_{\boldsymbol{p},\lambda'}g_{\boldsymbol{k},\lambda}
e^{-i(\boldsymbol{p}\cdot\boldsymbol{R}_i
-\boldsymbol{k}\cdot\boldsymbol{R}_j)}\mathcal G_{ij}(\omega_{\boldsymbol{k}}).
\label{eq:T_SM}
\end{align}
where the usual infinitesimal prescription $+i0^+$ has been omitted because radiatively coupled eigenmodes of $\hat H_{\mathrm{eff}}^{(1)}$ already have negative imaginary parts, while perfectly dark eigenmodes do not contribute to the scattering amplitude. For compactness, we have defined the single-excitation Green function
\begin{equation}
\mathcal G_{ij}(\omega)
\equiv
\left\langle G\right|
\hat\sigma_i
\frac{1}{\omega-\hat H_{\rm eff}^{(1)}}
\hat\sigma_j^\dagger
\left|G\right\rangle .
\label{eq:Gij2SM}
\end{equation}
Equations~\eqref{eq:A1SM} and \eqref{eq:T_SM} are the Markovian one-photon $S$ matrix for the full three-dimensional free-space continuum.  Because the original atom--field interaction is time independent, energy is conserved, while transverse momentum is not generally conserved by a finite array.  The latter enters through the phase factors for the actual $\boldsymbol R_j$.  Specular and diffuse channels are obtained only after projecting the resulting amplitude onto the desired far-field optical modes. For a single two-level atom, the Green function reduces to $\mathcal G(\omega)=1/(\omega-\omega_0+i\gamma_0/2)$, yielding a Lorentzian scattering spectrum.

\subsection{Two-photon scattering amplitude}
\label{sec:twoPhotonAmplitudeSM}
The scattering amplitude for two incoming photons $(\boldsymbol k_1,\lambda_1)$ and $(\boldsymbol k_2,\lambda_2)$ and two outgoing photons $(\boldsymbol k_3,\lambda_3)$ and $(\boldsymbol k_4,\lambda_4)$ is defined as
\begin{align}
\mathcal A^{(2)}(\boldsymbol k_3,\lambda_3,
\boldsymbol k_4,\lambda_4;
\boldsymbol k_1,\lambda_1,
\boldsymbol k_2,\lambda_2) 
&=\langle G,0|
\hat a_{\boldsymbol k_4,\lambda_4}
\hat a_{\boldsymbol k_3,\lambda_3}
\hat S
\hat a^\dagger_{\boldsymbol k_2,\lambda_2}
\hat a^\dagger_{\boldsymbol k_1,\lambda_1}
|0,G\rangle =\langle G,0|\hat{a}_{\boldsymbol{k_4},\lambda_4}^{\rm out}\hat{a}_{\boldsymbol{k_3},\lambda_3}^{\rm out}\hat{a}_{\boldsymbol{k_2},\lambda_2}^{\rm in\dagger}\hat{a}_{\boldsymbol{k_1},\lambda_1}^{\rm in\dagger}|0,G\rangle \\
=& \left\langle G,0\right|\left(\hat{a}_{\boldsymbol{k_4},\lambda_4}^{\rm in}+\hat{\mathcal S}_{\boldsymbol{k_4},\lambda_4}\right)\left(\hat{a}_{\boldsymbol{k_3},\lambda_3}^{\rm in}+\hat{\mathcal S}_{\boldsymbol{k_3},\lambda_3}\right) \hat{a}_{\boldsymbol{k_2},\lambda_2}^{\rm in\dagger}\hat{a}_{\boldsymbol{k_1},\lambda_1}^{\rm in\dagger}\left|0,G\right\rangle \equiv \mathcal B_1 + \mathcal B_2+\mathcal B_3+\mathcal B_4,
\end{align}
where the four terms are
\begin{align}
\mathcal B_1 & =  \left\langle G,0\right|\hat{a}_{\boldsymbol{k_4},\lambda_4}^{\rm in}\hat{a}_{\boldsymbol{k_3},\lambda_3}^{\rm in} \hat{a}_{\boldsymbol{k_2},\lambda_2}^{\rm in\dagger}\hat{a}_{\boldsymbol{k_1},\lambda_1}^{\rm in\dagger}\left|0,G\right\rangle,\qquad  \mathcal B_2 =  \left\langle G,0\right|\hat{\mathcal S}_{\boldsymbol{k_4},\lambda_4}\hat{a}_{\boldsymbol{k_3},\lambda_3}^{\rm in} \hat{a}_{\boldsymbol{k_2},\lambda_2}^{\rm in\dagger}\hat{a}_{\boldsymbol{k_1},\lambda_1}^{\rm in\dagger}\left|0,G\right\rangle, \\
\mathcal B_3 & =  \left\langle G,0\right| \hat{a}_{\boldsymbol{k_4},\lambda_4}^{\rm in}\hat{\mathcal S}_{\boldsymbol{k_3},\lambda_3}\hat{a}_{\boldsymbol{k_2},\lambda_2}^{\rm in\dagger}\hat{a}_{\boldsymbol{k_1},\lambda_1}^{\rm in\dagger}\left|0,G\right\rangle,\qquad  \mathcal B_4 =  \left\langle G,0\right|\hat{\mathcal S}_{\boldsymbol{k_4},\lambda_4}\hat{\mathcal S}_{\boldsymbol{k_3},\lambda_3} \hat{a}_{\boldsymbol{k_2},\lambda_2}^{\rm in\dagger}\hat{a}_{\boldsymbol{k_1},\lambda_1}^{\rm in\dagger}\left|0,G\right\rangle,
\end{align}

The free-field term yields the identity contribution
\begin{equation}
 \mathcal{B}_1 = (1+P_{34})\delta_{\lambda_1\lambda_3}\delta_{\lambda_2\lambda_4}\delta(\boldsymbol{k}_1-\boldsymbol{k}_3)\delta(\boldsymbol{k}_2-\boldsymbol{k}_4)  
\end{equation}
where $P_{mn}$ exchanges the labels $m$ and $n$. The term containing a single source operator reproduce the single-photon scattering coefficient $\mathcal T$ defined in Eq.~\eqref{eq:T_SM}
\begin{equation}
\mathcal{B}_2 = -2\pi i(1+P_{12})\mathcal T(\boldsymbol{k}_4,\lambda_4;
\boldsymbol{k}_2,\lambda_2)
\delta_{\lambda_1\lambda_3}\delta(\boldsymbol{k}_1-\boldsymbol{k}_3)\delta(\omega_2-\omega_4),   
\end{equation}
where $\omega_i=\omega_{\boldsymbol{k}_i}$. For $\mathcal{B}_3$, the incoming annihilation operator must first be commuted past the source operator: 
\begin{align}
\mathcal{B}_3 & = \left\langle G,0\right|\left(\hat{\mathcal S}_{\boldsymbol{k_3},\lambda_3}\hat{a}_{\boldsymbol{k_4},\lambda_4}^{\rm in}+\left[\hat{a}_{\boldsymbol{k_4},\lambda_4}^{\rm in},\hat{\mathcal S}_{\boldsymbol{k_3},\lambda_3}\right]\right) \hat{a}_{\boldsymbol{k_2},\lambda_2}^{\rm in\dagger}\hat{a}_{\boldsymbol{k_1},\lambda_1}^{\rm in\dagger}\left|0,G\right\rangle \\
& = -2\pi i(1+P_{12})\mathcal T(\boldsymbol{k}_3,\lambda_3;
\boldsymbol{k}_1,\lambda_1)
\delta_{\lambda_2\lambda_4}\delta(\boldsymbol{k}_2-\boldsymbol{k}_4)\delta(\omega_1-\omega_3) + \mathcal{B}_{3,\rm ex}.
\end{align}
The additional contribution is
\begin{equation}
\mathcal{B}_{3,\rm ex} = \left\langle G,0\right|\left[\hat{a}_{\boldsymbol{k_4},\lambda_4}^{\rm in},\hat{\mathcal S}_{\boldsymbol{k_3},\lambda_3}\right] \hat{a}_{\boldsymbol{k_2},\lambda_2}^{\rm in\dagger}\hat{a}_{\boldsymbol{k_1},\lambda_1}^{\rm in\dagger}\left|0,G\right\rangle = \left\langle G,0\right|\left[\left[[\hat{a}_{\boldsymbol{k_4},\lambda_4}^{\rm in},\hat{\mathcal S}_{\boldsymbol{k_3},\lambda_3}] ,\hat{a}_{\boldsymbol{k_2},\lambda_2}^{\rm in\dagger}\right],\hat{a}_{\boldsymbol{k_1},\lambda_1}^{\rm in\dagger}\right]\left|0,G\right\rangle.    
\end{equation}
Using the causal commutator \eqref{eq:CR_ain}, we find
\begin{equation}
\left[\hat{a}_{\boldsymbol{k_4},\lambda_4}^{\rm in},\hat{\mathcal S}_{\boldsymbol{k_3},\lambda_3}\right] =\sum_{j_3j_4}g^*_{\boldsymbol{k_3},\lambda_3}g^*_{\boldsymbol{k_4},\lambda_4} e^{-i(\boldsymbol{k}_3\cdot\boldsymbol{R}_{j_3}+\boldsymbol{k}_4\cdot\boldsymbol{R}_{j_4})}\int_{t_i}^{t_f}dt_3\int_{t_i}^{t_3}dt_4e^{i(\omega_3t_3+\omega_4t_4})\left[\hat{\sigma}_{j_4}(t_4),\hat{\sigma}_{j_3}(t_3)\right]. \label{eq:sigma4sigma3}
\end{equation}

We next rewrite the two-source contribution as nested commutator,
\begin{align}
 \mathcal{B}_4 & = \left\langle G,0\right|\left[[\hat{\mathcal S}_{\boldsymbol{k_4},\lambda_4}\hat{\mathcal S}_{\boldsymbol{k_3},\lambda_3},\hat{a}_{\boldsymbol{k_2},\lambda_2}^{\rm in \dagger}],\hat{a}_{\boldsymbol{k_1},\lambda_1}^{\rm in\dagger}\right]\left|0,G\right\rangle\\
 & = -\sum_{j_3j_4}g^*_{\boldsymbol{k_3},\lambda_3}g^*_{\boldsymbol{k_4},\lambda_4} e^{-i(\boldsymbol{k}_3\cdot\boldsymbol{R}_{j_3}+\boldsymbol{k}_4\cdot\boldsymbol{R}_{j_4})}\int_{t_i}^{t_f}dt_3\int_{t_i}^{t_f}dt_4e^{i(\omega_3t_3+\omega_4t_4})\langle G,0|\left[ [\hat{\sigma}_{j_4}(t_4)\hat{\sigma}_{j_3}(t_3),\hat{a}_{\boldsymbol{k_2},\lambda_2}^{\rm in\dagger}],\hat{a}_{\boldsymbol{k_1},\lambda_1}^{\rm in\dagger}\right]|0,G\rangle
\end{align}
To combine $\mathcal{B}_4$ with $\mathcal{B}_{3,\rm ex}$, we divide the integration domain into $t_4>t_3$ and $t_3>t_4$. The commutator in Eq.~\eqref{eq:sigma4sigma3} contributes only when $t_3>t_4$. In that domain, the operator identity 
\begin{equation*}
-\hat{\sigma}_{j_4}(t_4)\hat{\sigma}_{j_3}(t_3)+[\hat{\sigma}_{j_4}(t_4),\hat{\sigma}_{j_3}(t_3)]= -\hat{\sigma}_{j_3}(t_3)\hat{\sigma}_{j_4}(t_4),     
\end{equation*}
reverses the order of the two atomic lowering operators. Consequently
\begin{equation}
\mathcal{B}_{3,\rm ex}+\mathcal{B}_{4} = -\sum_{j_3j_4}g^*_{\boldsymbol{k_3},\lambda_3}g^*_{\boldsymbol{k_4},\lambda_4} e^{-i(\boldsymbol{k}_3\cdot\boldsymbol{R}_{j_3}+\boldsymbol{k}_4\cdot\boldsymbol{R}_{j_4})}\int_{t_i}^{t_f}dt_3\int_{t_i}^{t_f}dt_4e^{i(\omega_3t_3+\omega_4t_4})\langle G,0|\left[ \left[\mathcal{T}_t\left\{\hat{\sigma}_{j_4}(t_4)\hat{\sigma}_{j_3}(t_3)\right\},\hat{a}_{\boldsymbol{k_2},\lambda_2}^{\rm in\dagger}\right],\hat{a}_{\boldsymbol{k_1},\lambda_1}^{\rm in\dagger}\right],\label{eq:B3exB41}   
\end{equation}
where $\mathcal{T}_t$ denotes time-ordering,
\begin{equation}
\mathcal{T}_t\left\{\hat{\sigma}_{j_4}(t_4)\hat{\sigma}_{j_3}(t_3)\right\} =  \Theta(t_4-t_3)\hat{\sigma}_{j_4}(t_4)\hat{\sigma}_{j_3}(t_3)+\Theta(t_3-t_4)\hat{\sigma}_{j_3}(t_3)\hat{\sigma}_{j_4}(t_4),
\end{equation}
with the step function $\Theta (t)$.

To simplify Eq.~\eqref{eq:B3exB41}, we first establish an operator identity for inserting an atomic raising operator into a time-ordered product. Consider a time-ordered sequence of Heisenberg-picture atomic operators, 
\begin{equation}
\hat{Q} = \hat{O}_n(t_n)\hat{O}_{n-1}(t_{n-1})\cdots\hat{O}_1(t_1),\qquad t_n>t_{n-1}>\cdots t_1,    
\end{equation}
where each $\hat{O}_r(t_r)$ is either $\hat{\sigma}_j(t_r)$ or $\hat{\sigma}^{\dagger}_j(t_r)$. Using $[AB,C] = A[B,C]+[A,C]B$ and the Hermitian conjugate of Eq.~\eqref{eq:CR_ain}, we obtain
\begin{equation}
\left[\hat{Q},\hat{a}^{\rm in\dagger}_{\boldsymbol{k},\lambda}\right]=-i\sum_j g_{\boldsymbol{k},\lambda}e^{i\boldsymbol{k}\cdot R_j}\int_{t_i}^{t_f}dse^{-i\omega_{\boldsymbol{k}}s}\sum_{r=1}^n\Theta(t_r-s)\hat{O}_n(t_n)\cdots\hat{O}_{r+1}(t_{r+1})\left[\hat{O}_r(t_r),\hat{\sigma}^{\dagger}_j(s)\right]\hat{O}_{r-1}(t_{r-1})\cdots\hat{O}_1(t_1).   \end{equation}
For a fixed $s$ satisfying $t_{q+1}>s>t_q$, only terms with $r=q+1,q+2,\cdots,n$ contribute. Defining $\hat{Q}_{>s}=\hat{O}_n(t_n)\cdots\hat{O}_{q+1}(t_{q+1})$ and $\hat{Q}_{<s}=\hat{O}_q(t_q)\cdots\hat{O}_{1}(t_{1})$, the commutator expansion telescopes to
\begin{align}
\sum_{r=q+1}^n\hat{O}_n(t_n)\cdots\hat{O}_{r+1}(t_{r+1})\left[\hat{O}_r(t_r),\hat{\sigma}^{\dagger}_j(s)\right]\hat{O}_{r-1}(t_{r-1})\cdots\hat{O}_1(t_1) = \hat{Q}_{>s}\sigma_j(s)\hat{Q}_{<s}-\hat{\sigma}^{\dagger}_j(s)\hat{Q}.
\end{align}
The first term on the right-hand side is precisely the time-ordered product obtained by inserting $\hat{\sigma}_j^{\dagger}(s)$ into $\hat{Q}$. Consequently,
\begin{equation}
\sum_{r=1}^n\Theta(t_r-s)\hat{O}_n(t_n)\cdots\hat{O}_{r+1}(t_{r+1})\left[\hat{O}_r(t_r),\hat{\sigma}^{\dagger}_j(s)\right]\hat{O}_{r-1}(t_{r-1})\cdots\hat{O}_1(t_1) = \mathcal{T}_t\left\{\hat{Q}\hat{\sigma}^{\dagger}_j(s)\right\}-\hat{\sigma}^{\dagger}_j(s)\hat{Q}.   
\end{equation}
Substituting this result into the preceding commutator gives the exact operator identity
\begin{equation}
\left[\hat{Q},\hat{a}^{\rm in\dagger}_{\boldsymbol{k},\lambda}\right]=-i\sum_j g_{\boldsymbol{k},\lambda}e^{i\boldsymbol{k}\cdot R_j}\int_{t_i}^{t_f}dse^{-i\omega_{\boldsymbol{k}}s}\left[\mathcal{T}_t\left\{\hat{Q}\hat{\sigma}^{\dagger}_j(s)\right\}-\hat{\sigma}^{\dagger}_j(s)\hat{Q}\right]. \label{eq:CR_Qain_c}  
\end{equation}

Apply the identity \eqref{eq:CR_Qain_c} successively to the two input-photon creation operators in Eq.~\eqref{eq:B3exB41}, we obtain 
\begin{align}
\mathcal{B}_{3,\rm ex}+\mathcal{B}_{4} = &  \sum_{j_1j_2j_3j_4}g_{\boldsymbol{k_1},\lambda_1}g_{\boldsymbol{k_2},\lambda_2}g^*_{\boldsymbol{k_3},\lambda_3}g^*_{\boldsymbol{k_4},\lambda_4} e^{i(\boldsymbol{k}_1\cdot\boldsymbol{R}_{j_1}+\boldsymbol{k}_2\cdot\boldsymbol{R}_{j_2}-\boldsymbol{k}_3\cdot\boldsymbol{R}_{j_3}-\boldsymbol{k}_4\cdot\boldsymbol{R}_{j_4})} \nonumber\\
& \int_{t_i}^{t_f}dt_4 \int_{t_i}^{t_f}dt_3\int_{t_i}^{t_f}dt_2\int_{t_i}^{t_f}dt_1\langle G,0|\mathcal{T}_t\left\{\hat{\sigma}_{j_4}(t_4)\hat{\sigma}_{j_3}(t_3)\hat{\sigma}^{\dagger}_{j_2}(t_2)\hat{\sigma}^{\dagger}_{j_1}(t_1)\right\}|0,G\rangle e^{i(\omega_4t_4+\omega_3t_3-\omega_2t_2-\omega_1t_1)}\\
= & \sum_{j_1j_2j_3j_4}g_{\boldsymbol{k_1},\lambda_1}g_{\boldsymbol{k_2},\lambda_2}g^*_{\boldsymbol{k_3},\lambda_3}g^*_{\boldsymbol{k_4},\lambda_4} e^{i(\boldsymbol{k}_1\cdot\boldsymbol{R}_{j_1}+\boldsymbol{k}_2\cdot\boldsymbol{R}_{j_2}-\boldsymbol{k}_3\cdot\boldsymbol{R}_{j_3}-\boldsymbol{k}_4\cdot\boldsymbol{R}_{j_4})} \sum_{m=1}^8\mathcal{C}_m,
\end{align}
The second term in the square brackets of Eq.~\eqref{eq:CR_Qain_c} vanishes due to $\langle G,0|\hat{\sigma}^{\dagger}_j(s)=0$. Equivalently, the four-time atomic correlation function can be obtained by expanding the scattering operator in the Dyson series and contracting the four external photon operators. The internal atom–field dynamics is encoded in the full Heisenberg-picture atomic operators, and both approaches lead to the same result. The eight contributions $\mathcal{C}_m$ correspond to the eight non-vanishing time orderings of the four atomic operators as summarized in the following table
\begin{center}\small
\begin{tabular}{ccl}
\toprule
symbol & time ordering & ordered operator-chain \\
\midrule
$\mathcal C_1$ & $t_4>t_3>t_2>t_1$ & $\langle\hat\sigma_{j_4}(t_4)\hat\sigma_{j_3}(t_3)\hat\sigma_{j_2}^\dagger(t_2)\hat\sigma_{j_1}^\dagger(t_1)\rangle$\\
$\mathcal C_2$ & $t_4>t_3>t_1>t_2$ & $\langle\hat\sigma_{j_4}(t_4)\hat\sigma_{j_3}(t_3)\hat\sigma_{j_1}^\dagger(t_1)\hat\sigma_{j_2}^\dagger(t_2)\rangle$\\
$\mathcal C_3$ & $t_3>t_4>t_2>t_1$ & $\langle\hat\sigma_{j_3}(t_3)\hat\sigma_{j_4}(t_4)\hat\sigma_{j_2}^\dagger(t_2)\hat\sigma_{j_1}^\dagger(t_1)\rangle$\\
$\mathcal C_4$ & $t_3>t_4>t_1>t_2$ & $\langle\hat\sigma_{j_3}(t_3)\hat\sigma_{j_4}(t_4)\hat\sigma_{j_1}^\dagger(t_1)\hat\sigma_{j_2}^\dagger(t_2)\rangle$\\
$\mathcal C_5$ & $t_4>t_2>t_3>t_1$ & $\langle\hat\sigma_{j_4}(t_4)\hat\sigma_{j_2}^\dagger(t_2)\hat\sigma_{j_3}(t_3)\hat\sigma_{j_1}^\dagger(t_1)\rangle$\\
$\mathcal C_6$ & $t_3>t_1>t_4>t_2$ & $\langle\hat\sigma_{j_3}(t_3)\hat\sigma_{j_1}^\dagger(t_1)\hat\sigma_{j_4}(t_4)\hat\sigma_{j_2}^\dagger(t_2)\rangle$\\
$\mathcal C_7$ & $t_3>t_2>t_4>t_1$ & $\langle\hat\sigma_{j_3}(t_3)\hat\sigma_{j_2}^\dagger(t_2)\hat\sigma_{j_4}(t_4)\hat\sigma_{j_1}^\dagger(t_1)\rangle$\\
$\mathcal C_8$ & $t_4>t_1>t_3>t_2$ & $\langle\hat\sigma_{j_4}(t_4)\hat\sigma_{j_1}^\dagger(t_1)\hat\sigma_{j_3}(t_3)\hat\sigma_{j_2}^\dagger(t_2)\rangle$\\
\bottomrule
\end{tabular}\label{table:ordering}
\end{center}

Using the quantum regression theorem, we evaluate the eight non-vanishing contributions. The first four orderings, in which both absorptions precede both emissions, give
\begin{equation}
\mathcal{C}_1+\mathcal{C}_2+\mathcal{C}_3+\mathcal{C}_4 = -2\pi i \mathcal M_{j_3j_4;j_1j_2}(\omega_3,\omega_4;\omega_1,\omega_2) \delta(\omega_3+\omega_4-\omega_1-\omega_2)  
\end{equation}
where 
\begin{equation}
\mathcal M_{j_{3}j_{4};j_{1}j_{2}}(\omega_{3},\omega_{4};\omega_{1},\omega_{2}) = (1+P_{34})(1+P_{12})\left\langle G\right|\hat{\sigma}_{j_4}
\frac{1}{\omega_4-\hat{H}^{(1)}_{{\rm eff}}}\hat{\sigma}_{j_3}
\frac{1}{\omega_1+\omega_2-\hat{H}^{(2)}_{{\rm eff}}}
\hat{\sigma}^{\dagger}_{j_2}\frac{1}{\omega_1-\hat{H}^{(1)}_{{\rm eff}}}\hat{\sigma}^{\dagger}_{j_1}\left|G\right\rangle.  \label{eq:M}  
\end{equation}
The operator $\hat{H}^{(2)}_{\rm eff}$ denotes the projection of the effective non-Hermitian Hamiltonian onto the two-excitation subspace.

Consider next the two alternating time orderings $t_4>t_2>t_3>t_1$ and $t_3>t_1>t_4>t_2$. Performing the ordered time integrals yields
\begin{align}
\mathcal C_5
&=
2\pi i \frac{
\mathcal G_{j_4j_2}(\omega_4)
\mathcal G_{j_3j_1}(\omega_1)
}{
\omega_3-\omega_1-i0^+
}\delta(\omega_3+\omega_4-\omega_1-\omega_2),\\
\mathcal C_6
&=
2\pi i \frac{
\mathcal G_{j_3j_1}(\omega_3)
\mathcal G_{j_4j_2}(\omega_2)
}{
\omega_4-\omega_2-i0^+
}\delta(\omega_3+\omega_4-\omega_1-\omega_2),
\label{eq:A2C56SM}
\end{align}
Energy conservation implies
$x=\omega_3-\omega_1
=-(\omega_4-\omega_2)$.
Using the Sokhotski–Plemelj identity
\begin{equation}
\frac{1}{x\mp i0^+} = \mathcal P\frac{1}{x}
\pm i\pi\delta(x),
\end{equation}
we obtain
\begin{align}
\mathcal C_5+\mathcal C_6
= -(2\pi)^2
\mathcal G_{j_4j_2}(\omega_2)
\mathcal G_{j_3j_1}(\omega_1)\delta(\omega_3-\omega_1)
\delta(\omega_4-\omega_2)+
2\pi i \,
\mathcal N^{(13|24)}_{j_3j_4;j_1j_2}\delta(\omega_3+\omega_4-\omega_1-\omega_2),
\label{eq:A2C56resultSM}
\end{align}
where
\begin{align}
\mathcal N^{(13|24)}_{j_3j_4;j_1j_2}(\omega_{3},\omega_{4};\omega_{1},\omega_{2})
=
\mathcal{P}\frac{1}{\omega_3-\omega_1}
\big[
\mathcal G_{j_4j_2}(\omega_4)
\mathcal G_{j_3j_1}(\omega_1)-
\mathcal G_{j_3j_1}(\omega_3)
\mathcal G_{j_4j_2}(\omega_2)
\big].
\label{eq:A2K13SM}
\end{align}
The first term in Eq.~\eqref{eq:A2C56resultSM} enforces independent energy conservation for the two photons and belongs to the disconnected amplitude. In contrast, the second term conserves only their total energy and contributes to the connected amplitude. The principal-value contributions do not cancel because they multiply different frequency-dependent single-excitation Green functions. The remaining alternating orderings,
$t_3>t_2>t_4>t_1$ and $t_4>t_1>t_3>t_2$, are evaluated in the same way. Their sum is
\begin{align}
\mathcal C_7+\mathcal C_8
= -(2\pi)^2
\mathcal G_{j_4j_1}(\omega_1)
\mathcal G_{j_3j_2}(\omega_2)\delta(\omega_4-\omega_1)
\delta(\omega_3-\omega_2) + 2\pi i
\mathcal N^{(23|14)}_{j_3j_4;j_1j_2} \delta(\omega_3+\omega_4-\omega_1-\omega_2),
\label{eq:A2C78resultSM}
\end{align}
with
\begin{equation}
\mathcal N^{(23|14)}_{j_3j_4;j_1j_2}(\omega_{3},\omega_{4};\omega_{1},\omega_{2})
= \mathcal{P}\frac{1}{\omega_4-\omega_1}\left[\mathcal G_{j_3j_2}(\omega_3)
\mathcal G_{j_4j_1}(\omega_1)-
\mathcal G_{j_4j_1}(\omega_4)
\mathcal G_{j_3j_2}(\omega_2)\right].\label{eq:A2K23SM}
\end{equation}

Combining all eight nonvanishing orderings with the direct and single-photon scattering contributions, we obtain
\begin{equation}
\mathcal A^{(2)}
=\mathcal A^{(2)}_{\rm dis}
+\mathcal A^{(2)}_{\rm con}.
\end{equation}
The disconnected amplitude describes independent scattering of the two photons:
\begin{align}
\mathcal A^{(2)}_{\rm dis}
= &
\mathcal A^{(1)}(\boldsymbol k_3,\lambda_3;
\boldsymbol k_1,\lambda_1)
\mathcal A^{(1)}(\boldsymbol k_4,\lambda_4;
\boldsymbol k_2,\lambda_2) +
\mathcal A^{(1)}(\boldsymbol k_4,\lambda_4;
\boldsymbol k_1,\lambda_1)
\mathcal A^{(1)}(\boldsymbol k_3,\lambda_3;
\boldsymbol k_2,\lambda_2).
\label{eq:A2disSM}
\end{align}
The connected scattering amplitude is 
\begin{align}
\mathcal A^{(2)}_{\rm con}=-2\pi i
\delta(\omega_3+\omega_4-\omega_1-\omega_2)
\sum_{j_1j_2j_3j_4}g^*_{\boldsymbol k_3,\lambda_3}
g^*_{\boldsymbol k_4,\lambda_4}
g_{\boldsymbol k_1,\lambda_1}
g_{\boldsymbol k_2,\lambda_2}e^{i(
\boldsymbol k_1\cdot\boldsymbol R_{j_1}
+\boldsymbol k_2\cdot\boldsymbol R_{j_2}
-\boldsymbol k_3\cdot\boldsymbol R_{j_3}
-\boldsymbol k_4\cdot\boldsymbol R_{j_4}
)} (\mathcal M_{j_3j_4;j_1j_2}-\mathcal N_{j_3j_4;j_1j_2}).
\label{eq:A2conSM}
\end{align}
where $\mathcal M_{j_3j_4;j_1j_2}$ is given in Eq.~\eqref{eq:M} and 
\begin{align}
\mathcal N_{j_3j_4;j_1j_2}(\omega_{3},\omega_{4};\omega_{1},\omega_{2}) & =  \mathcal N^{(13|24)}_{j_3j_4;j_1j_2}(\omega_{3},\omega_{4};\omega_{1},\omega_{2}) + \mathcal N^{(23|14)}_{j_3j_4;j_1j_2}(\omega_{3},\omega_{4};\omega_{1},\omega_{2})\\
& = (1+P_{34})\mathcal P\frac{1}{\omega_3-\omega_1}\left[\mathcal{G}_{j_4j_2}(\omega_4)\mathcal{G}_{j_3j_1}(\omega_1)-\mathcal{G}_{j_3j_1}(\omega_3)\mathcal{G}_{j_4j_2}(\omega_2)\right].   
\end{align}
The four terms in $\mathcal M$ describe propagation through the two-excitation subspace, whereas the two terms in $\mathcal N$ arise from alternating
absorption–emission orderings. Both are required to construct the complete connected two-photon amplitude. In particular, for a single two-level atom the two-excitation subspace is absent, so $\mathcal M= 0$, but $\mathcal N$ remains nonzero and captures the nonlinear response arising from saturation.

\section{Specular scattering of a paraxial Gaussian single-photon pulse}

The incident one-photon pulse is written as
\begin{equation}
|1_{\rm in}\rangle=
\int d\boldsymbol{q}\int_0^\infty \frac{d\omega}{\sqrt{v_z}}
A(\boldsymbol{q},\omega)\hat a^\dagger_{\boldsymbol{q},-k_z,-}|0\rangle,
\qquad
k_z=\sqrt{\omega^2/c^2-q^2},
\label{eq:1inSM}
\end{equation}
where the incident photon propagates along the negative $z$ direction, $\boldsymbol{q}=\{k_x,k_y\}$ is the transverse wave vector, and $v_{z}=|\partial\omega_{\boldsymbol{k}}/\partial k_{z}|\approx c$. We use a quasi-monochromatic paraxial spectrum
\begin{equation}
A(\boldsymbol{q},\omega)=\xi(\boldsymbol{q})\chi(\omega),
\qquad
\int d\boldsymbol{q}\,|\xi(\boldsymbol{q})|^2=\int d\omega |\chi(\omega)|^2=1,
\end{equation}
with $\chi(\omega)$ centered at $\omega_{\rm in}$ and with bandwidth much smaller than $\gamma_0$.  For a Gaussian transverse mode
\begin{equation}
\xi(\boldsymbol{q})=\frac{w_0}{\sqrt{2\pi}}\exp\left(-\frac{w_0^2q^2}{4}\right),
\label{eq:xiSM}
\end{equation}
the corresponding amplitude sampled at the atomic sites is
\begin{equation}
\psi_j=\frac{1}{2\pi}\int d\boldsymbol{q} \xi(\boldsymbol{q})\exp^{i\boldsymbol{q}\cdot\boldsymbol{R}_j}
=\sqrt{\frac{2}{\pi w_0^2}}\exp\left[-\frac{x_j^2+y_j^2}{w_0^2}\right].
\label{eq:psiSM}
\end{equation}

The ideal specular transmitted and reflected one-photon states are
\begin{align}
|1_T\rangle&=\int d\boldsymbol{q}\int_0^\infty \frac{d\omega}{\sqrt{v_z}} 
A(\boldsymbol{q},\omega)\hat a^\dagger_{\boldsymbol{q},-k_z,-}|0\rangle,\\
|1_R\rangle&=\int d\boldsymbol{q} \int_0^{\infty}\frac{d\omega}{\sqrt{v_z}} A(\boldsymbol{q},\omega)\hat a^\dagger_{\boldsymbol{q},+k_z,+}|0\rangle.
\end{align}
The specular amplitudes are
\begin{equation}
r_{\rm spec}=\langle 1_R|\hat S|1_{\rm in}\rangle,
\qquad
t_{\rm spec}=\langle 1_T|\hat S|1_{\rm in}\rangle .
\end{equation}
In the paraxial limit, the on-shell delta function in Eq.~\eqref{eq:A1SM} is evaluated as
\begin{equation}
\delta(\omega_{\boldsymbol{p}}-\omega_{\boldsymbol{k}})
=\frac{1}{v_z}\delta(p_z -s_\alpha k_z)
\simeq \frac{1}{c}\delta(p_z-s_\alpha k_z),
\label{eq:deltaSM}
\end{equation}
where $\alpha = R,T$ denotes reflection and transmission, $s_T=-1$, and $s_R=1$. 

The
paraxial polarization overlaps and the slowly varying couplings may be replaced by
\begin{equation*}
g^*_{\boldsymbol k_\alpha,\lambda_\alpha}
 g_{\boldsymbol k',-}
\simeq - s_\alpha|g_0|^2,\qquad
|g_0|^2=\frac{\omega_0|d|^2}{2\hbar\varepsilon_0(2\pi)^3}.
\end{equation*}
In the smooth paraxial helicity convention, the incident and detected polarization vectors approach $\boldsymbol e_{\boldsymbol k',-}\simeq-\boldsymbol e_+$, $\boldsymbol e_{\boldsymbol k_R,+}\simeq\boldsymbol e_+$, and $\boldsymbol e_{\boldsymbol k_T,-}\simeq-\boldsymbol e_+$,
which gives the stated relative sign. Since the atoms lie in the $xy$ plane, the phase of the outgoing wave contributes
$e^{-i\boldsymbol q\cdot\boldsymbol R_i}$, while the integral over the incident transverse momenta gives
\begin{equation*}
\int d^2q'\,\xi(\boldsymbol q')
 e^{i\boldsymbol q'\cdot\boldsymbol R_j}
=2\pi\psi_j.
\end{equation*} 
Finally, we obtain
\begin{align}
t_{\rm spec}
&=1-i\frac{3\pi\gamma_0}{2k_0^2}
\sum_{ij}\psi_i^*
\left\langle G\right|
\hat\sigma_i\frac{1}{\omega_{\rm in}-\hat H^{(1)}_{\rm eff}}
\hat\sigma_j^\dagger\left|G\right\rangle\psi_j,
\label{eq:tspecSM}\\
r_{\rm spec}
&=i\frac{3\pi\gamma_0}{2k_0^2}
\sum_{ij}\psi_i^*
\left\langle G\right|
\hat\sigma_i\frac{1}{\omega_{\rm in}-\hat H^{(1)}_{\rm eff}}
\hat\sigma_j^\dagger\left|G\right\rangle\psi_j,
\label{eq:rspecSM}
\end{align}
where $k_0=\omega_{\rm in}/c$. The specular rates are $T_{\rm spec}=|t_{\rm spec}|^2$ and $R_{\rm spec}=|r_{\rm spec}|^2$.  The remaining probability is diffuse scattering into non-specular free-space modes,
\begin{equation}
D_{\rm dif}=1-R_{\rm spec}-T_{\rm spec},
\end{equation}
when nonradiative decay is absent.

\section{Collective-mode expansion and Rayleigh threshold}

Let the right and left eigenvectors of $\hat H^{(1)}_{\rm eff}$ satisfy
\begin{equation}
\hat H^{(1)}_{\rm eff}|n_R\rangle=E_n|n_R\rangle,
\qquad
\langle n_L|\hat H^{(1)}_{\rm eff}=E_n\langle n_L|,
\qquad
\langle n_L|m_R\rangle=\delta_{nm}.
\end{equation}
For the reciprocal Green-tensor coupling considered here, the single-excitation matrix is complex symmetric.  Thus the left eigenvectors are the transpose, not the Hermitian conjugate, of the right eigenvectors.  If $U$ is the matrix whose columns are $|n_R\rangle$, then $U^T U=I$, and
\begin{equation}
\left\langle G\right|
\hat\sigma_i\frac{1}{\omega_{\rm in}-\hat H^{(1)}_{\rm eff}}
\hat\sigma_j^\dagger\left|G\right\rangle
=\left[U\frac{1}{\omega_{\rm in}-\Lambda}U^T\right]_{ij},
\end{equation}
where $\Lambda={\rm diag}(E_1,E_2,\ldots)$.  Writing
\begin{equation}
E_n=\omega_n-i\frac{\Gamma_n}{2},
\qquad
\Delta_n=\omega_{\rm in}-\omega_n,
\end{equation}
Eqs.~\eqref{eq:tspecSM} and \eqref{eq:rspecSM} reduce to
\begin{align}
r_{\rm spec} & =i\frac{3\pi\gamma_0}{2k_0^2}
\sum_n \frac{B_n}{\Delta_n+i\Gamma_n/2},
\label{eq:rspecmodeSM}\\
t_{\rm spec} & =1-i\frac{3\pi\gamma_0}{2k_0^2}
\sum_n \frac{B_n}{\Delta_n+i\Gamma_n/2},
\label{eq:tspecmodeSM}
\end{align}
with the complex residue
\begin{equation}
B_n=\sum_{ij}\psi_i^*U_{in}U_{jn}\psi_j .
\label{eq:BnSM}
\end{equation}

\begin{figure}
\includegraphics[width=8.5cm]{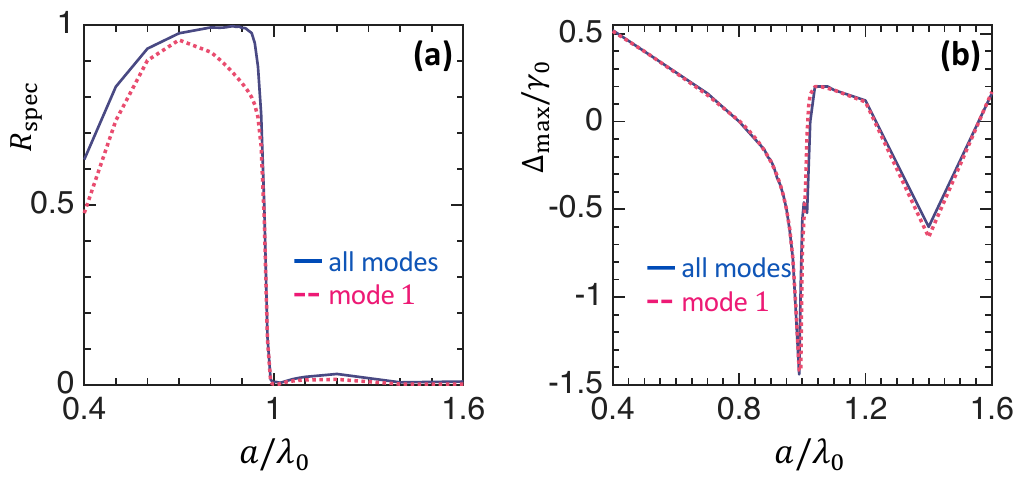}
\centering
\caption{Dominant-mode description of the specular reflection. As indicated by Eq.~(\ref{eq:rspecmodeSM}), the specular reflection amplitude can be decomposed into contributions from all collective eigenmodes in the single-excitation subspace. (a) Maximum specular reflection $R_{\rm spec}$ obtained by retaining all modes (solid line) and only the dominant mode with the largest residue (dashed line). (b) Detuning $\Delta_{\rm max}$ at which the full reflection spectrum reaches its Lorentzian maximum, compared with the resonance position predicted from the eigenenergy of the dominant mode. The close agreement shows that the specular response is governed primarily by a single collective mode over most of the parameter range. The parameters are $L=40$ and $w_0=10\lambda_0$. \label{fig:1mode_approximation}}
\end{figure}

Figure~\ref{fig:1mode_approximation} shows that the specular response is dominated by a single collective mode over nearly the entire parameter range. As seen in panel (a), retaining only the mode with the largest residue closely reproduces the full multimode result, including the near-unity reflection below the Rayleigh threshold and its abrupt collapse around $a_{\rm RT}=\lambda_0$. Noticeable deviations occur only within a narrow interval immediately below the threshold, where additional modes make a non-negligible contribution as explained in main text. Panel (b) shows similarly close agreement in the resonance position: the detuning $\Delta_{\rm max}$ extracted from the full reflection spectrum nearly coincides with that predicted by the eigenenergy of the dominant mode. These results confirm that, except in the small pre-threshold region, both the magnitude and spectral position of the specular reflection are governed primarily by a single collective excitation.

For an infinite square lattice, the single-excitation eigenstates are Bloch spin waves
\begin{equation}
|\boldsymbol{q}\rangle=\frac{1}{\sqrt N}\sum_j\exp^{i\boldsymbol{q}\cdot\boldsymbol{R}_j}\hat\sigma_j^\dagger|G\rangle .
\end{equation}
Their radiative linewidth can be written as~\cite{AsenjoGarcia2017Selective}
\begin{equation}
\frac{\Gamma_{\boldsymbol{q}}}{\gamma_0}
=\frac{3\pi}{k_0^3a^2}
\sum_{\boldsymbol g\in|\boldsymbol{q}+\boldsymbol g|\le k_0}
\frac{k_0^2-| (\boldsymbol{q}|+\boldsymbol{g})\cdot\boldsymbol d/d|^2}
{\sqrt{k_0^2-|\boldsymbol{q}+\boldsymbol{g}|^2}},
\label{eq:GammaqSM}
\end{equation}
where $\boldsymbol{g}=(2\pi/a)(m\boldsymbol{e}_x+n\boldsymbol{e}_y)$ is a reciprocal lattice vector.  For the near-normal spin waves addressed by the paraxial beam, $\boldsymbol{q}\simeq0$.  Below the first diffraction threshold, only the zeroth order $\boldsymbol{g}=0$ radiates.  At the Rayleigh threshold $a_{\rm RT}$, the first diffraction orders $(m,n)=(\pm1,0),(0,\pm1)$ reach the light cone, $|\boldsymbol{q}+\boldsymbol{g}|\simeq k_0$, so the longitudinal wave number of the emitted photon vanishes and the denominator in Eq.~\eqref{eq:GammaqSM} becomes singular.  This is the Rayleigh anomaly.  For a finite array driven by a finite-waist beam, the singularity is rounded into a sharp but finite linewidth peak, which redirects radiation into near-grazing modes with $k_z\simeq0$ and suppresses specular reflection.  The next set of reciprocal vectors, $(m,n)=(\pm1,\pm1)$, reaches the light cone near $a=\sqrt2\lambda_0$ and produces the weaker second feature discussed in the main text.

\begin{figure}
\includegraphics[width=15cm]{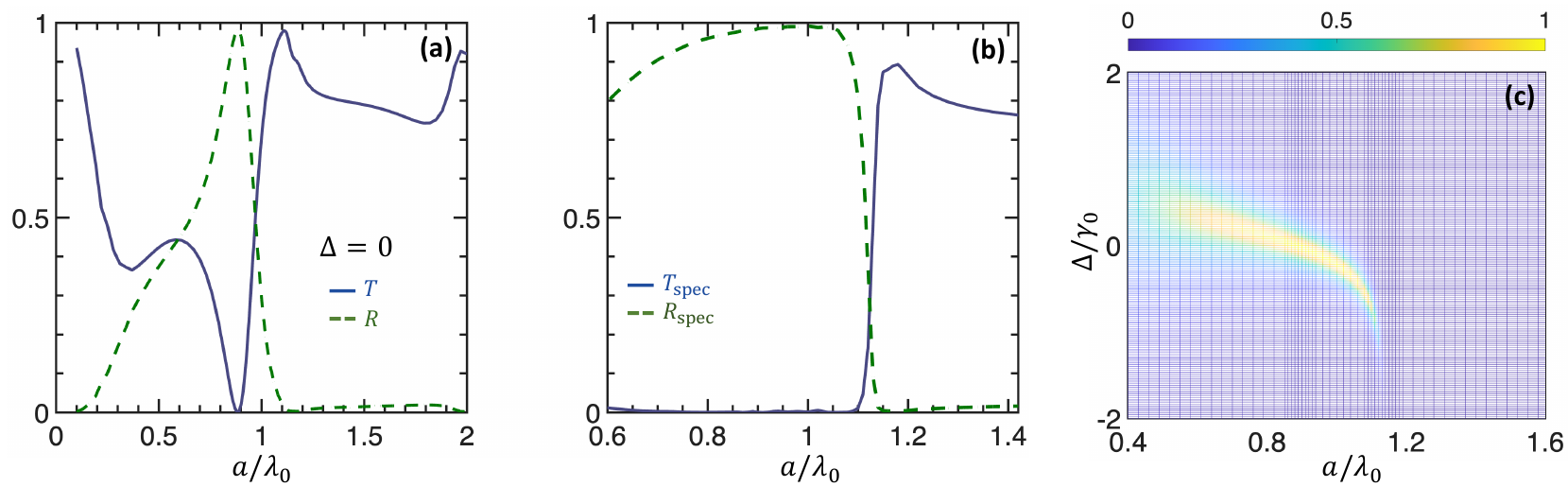}
\centering
\caption{Distinguishing fixed-frequency extinction from the Rayleigh threshold in a finite triangular atomic array. (a) Specular transmission and reflection at fixed detuning ($\Delta=0$), showing the strong lattice-spacing dependence previously reported in Ref.~\cite{Bettles2016Enhanced}. (b) Frequency-optimized specular reflection and corresponding transmission, revealing an abrupt collapse of reflection near the first Rayleigh threshold ($a_{\mathrm{RW}}=2\lambda_0/\sqrt{3}$). (c) Specular reflection spectrum as a function of lattice spacing and detuning, demonstrating that the loss of high reflectivity near the threshold cannot be recovered by tuning the incident frequency.\label{fig:triangular}}
\end{figure}

Figure~\ref{fig:triangular} further establishes that the Rayleigh threshold is fundamentally a geometric effect, distinct from the magic lattice spacing identified in Ref.~\cite{Bettles2016Enhanced}. The latter corresponds to enhanced extinction arising from cooperative interference at a particular incident detuning, whereas the Rayleigh threshold is determined solely by the lattice geometry and the opening of additional radiative diffraction channels. For a triangular lattice under normal incidence, this threshold occurs at $a_{\rm RT}=2\lambda_0/\sqrt{3}$, in contrast to $a_{\rm RT}=\lambda_0$ for a square lattice. Importantly, Fig.~\ref{fig:triangular}(c) demonstrates that the collapse of specular reflection persists even after optimizing the incident frequency, confirming that it cannot be attributed to a particular choice of detuning. The agreement between square and triangular lattices establishes the Rayleigh threshold as a universal geometric mechanism for controlling collective optical scattering in two-dimensional periodic atomic arrays, with the precise threshold determined by the reciprocal-lattice geometry.

\section{Weak coherent input and normalized correlation function}
For a weak coherent-state input in the same spatiotemporal mode as Eq.~\eqref{eq:1inSM},
\begin{equation}
|\alpha_{\rm in}\rangle
=|0\rangle+\alpha|1_{\rm in}\rangle
+\frac{\alpha^2}{\sqrt2}|2_{\rm in}\rangle+O(\alpha^3),\label{eq:coherentstate}
\end{equation}
where
\begin{equation}
|2_{\rm in}\rangle=\frac{1}{\sqrt2}
\left[\int d\boldsymbol{q}\int_0^\infty d\omega 
A(\boldsymbol{q},\omega)
\hat a^\dagger_{\boldsymbol{q},-k_z,-}\right]^2|0\rangle .
\end{equation}
To leading order in $|\alpha|$, the single-photon intensity is determined by the outgoing one-photon state $|\Psi^{(1)}_{\rm out}\rangle$, while the normally ordered second-order correlation is determined by $|\Psi^{(2)}_{\rm out}\rangle$.  Thus
\begin{equation}
I_\alpha(\boldsymbol{\rho})
=\langle\Psi^{(1)}_{\rm out}|\hat E_\alpha^{(-)}(\boldsymbol{\rho})
\hat E_\alpha^{(+)}(\boldsymbol{\rho})|\Psi^{(1)}_{\rm out}\rangle
\equiv |\mathcal A^{(1)}_\alpha(\boldsymbol{\rho})|^2,
\end{equation}
and
\begin{align}
G^{(2)}_{\alpha\alpha}(\boldsymbol{\rho},\boldsymbol{\rho'})
&=\langle\Psi^{(2)}_{\rm out}|
\hat E_\alpha^{(-)}(\boldsymbol{\rho})
\hat E_\alpha^{(-)}(\boldsymbol{\rho'})
\hat E_\alpha^{(+)}(\boldsymbol{\rho'})
\hat E_\alpha^{(+)}(\boldsymbol{\rho})
|\Psi^{(2)}_{\rm out}\rangle=\left|\langle0|\hat E_\alpha^{(+)}(\boldsymbol{\rho'})
\hat E_\alpha^{(+)}(\boldsymbol{\rho})|\Psi^{(2)}_{\rm out}\rangle\right|^2
\equiv |\mathcal A^{(2)}_{\alpha\alpha}(\boldsymbol{\rho},\boldsymbol{\rho'})|^2 .
\end{align}
The normalized coherence function is defined as
\begin{equation}
g^{(2)}_{\alpha\alpha}(\boldsymbol{\rho},\boldsymbol{\rho'})
=\frac{G^{(2)}_{\alpha\alpha}(\boldsymbol{\rho},\boldsymbol{\rho'})}
{2I_\alpha(\boldsymbol{\rho})I_\alpha(\boldsymbol{\rho'})}.
\label{eq:g2defSM}
\end{equation}
The factor of $2$ comes from the normalization of the two-photon Fock component in the weak coherent-state expansion.

\section{Momentum-space detection with a 2$f$ system}
A $2f$ lens system maps the transverse wave vector of a collected photon
onto the camera coordinate,
\begin{equation}
\boldsymbol q=\frac{k_0}{f}\boldsymbol\rho,
\label{eq:2fmapSM}
\end{equation}
where $f$ is the focal length. This mapping is understood in the quasi-monochromatic, paraxial approximation. The finite collection aperture is described by
\begin{equation}
P(\boldsymbol q)=
\begin{cases}
1, & |\boldsymbol q|<q_c,\\
0, & |\boldsymbol q|\ge q_c,
\end{cases}
\qquad q_c=k_0\sin\theta_c\simeq k_0\theta_c.
\end{equation}
In our calculations, we choose $q_c=0.16k_0\simeq1/\lambda_0$, corresponding to $\mathrm{NA}=\sin\theta_c=0.16$ ($\theta_c\simeq9.2^\circ$), so that only near-axis modes are collected and the scalar low-NA approximation employed below is justified. In the scalar low-numerical-aperture (NA) approximation, the detected positive-frequency field is
\begin{equation}
\hat E^{(+)}_\alpha(\boldsymbol\rho)
\simeq \tilde C_\alpha\int_0^\infty d\omega\,
P(\boldsymbol q)e^{i\phi_\alpha(k_z)}
\hat a_{\boldsymbol k_\alpha,\lambda_\alpha},
\qquad
\boldsymbol k_\alpha=(\boldsymbol q,s_\alpha k_z),
\quad k_z=\sqrt{\omega^2/c^2-q^2},
\label{eq:E2fSM}
\end{equation}
where $\alpha=R,T$ denotes reflection and transmission, $s_R=+1$ and $s_T=-1$. The outgoing polarizations $\lambda_R=+$ and $\lambda_T=-$ are defined in the same helicity convention as for the incident field. The constant $\tilde C_\alpha$ contains the field and lens normalization.

\subsection{Momentum-resolved single-photon intensity}
A $2f$ lens system performs a spatial Fourier transform, mapping the transverse momentum distribution of the scattered light onto the intensity distribution in its back focal plane via the relation in Eq.~\eqref{eq:2fmapSM}. To obtain the one-photon detection amplitude, we project the plane-wave $S$ matrix onto the incident Gaussian wave packet,
\begin{align*}
\mathcal A^{(1)}_\alpha(\boldsymbol\rho)
&\equiv\langle G,0|\hat E^{(+)}_\alpha(\boldsymbol\rho)
\hat S|1_{\rm in}\rangle\\
&=\tilde C_\alpha P(\boldsymbol q)
\int_0^\infty d\omega e^{i\phi_\alpha(k_z)}
\int d\boldsymbol{q}'\int_0^\infty\frac{d\omega'}{\sqrt{v_z'}}\,
\xi(\boldsymbol q')\chi(\omega')
\mathcal A^{(1)}(\boldsymbol k_\alpha,\lambda_\alpha;
\boldsymbol k',-),
\end{align*}
where $\boldsymbol k'=(\boldsymbol q',-k_z')$,
$k_z'=\sqrt{\omega'^2/c^2-q'^2}$, and
$v_z'=|\partial\omega'/\partial k_z'|\simeq c$. Inserting the plane-wave one-photon amplitude~\eqref{eq:A1SM}, the on-shell factor $\delta(\omega-\omega')$ fixes the output frequency.  An alternative overall phase convention for the reflected mode changes its amplitude but not the
measured intensity or normalized correlation. Neglecting the frequency dependence of $\mathcal G_{ij}(\omega)$, the
polarization overlaps, $v_z$, and the lens phase across the incident bandwidth, we obtain
\begin{align}
\mathcal A^{(1)}_R(\boldsymbol{\rho})
&=i P(\boldsymbol{q})C_R\frac{3\gamma_0}{4k_0^2}
\sum_{ij}\exp^{-i\boldsymbol{q}\cdot\boldsymbol{R}_i}
\left\langle G\right|
\hat\sigma_i\frac{1}{\omega_{\rm in}-\hat H^{(1)}_{\rm eff}}
\hat\sigma_j^\dagger\left|G\right\rangle\psi_j,
\label{eq:AR2fSM}\\
\mathcal A^{(1)}_T(\boldsymbol{\rho})
&=P(\boldsymbol{q})C_T\left[
\xi(\boldsymbol{q})-i\frac{3\gamma_0}{4k_0^2}
\sum_{ij}\exp^{-i\boldsymbol{q}\cdot\boldsymbol{R}_i}
\left\langle G\right|
\hat\sigma_i\frac{1}{\omega_{\rm in}-\hat H^{(1)}_{\rm eff}}
\hat\sigma_j^\dagger\left|G\right\rangle\psi_j
\right],
\label{eq:AT2fSM}
\end{align}
where 
\begin{equation}
C_\alpha \equiv \sqrt c \tilde C_\alpha e^{i\phi_\alpha(\boldsymbol q)}
 \int_0^\infty d\omega\,\chi(\omega). \label{eq:Calpha}   
\end{equation}
The directly transmitted Gaussian amplitude $\xi(\boldsymbol q)$ interferes with the field coherently scattered by the array. Their destructive interference produces the momentum-space Fano intensity zeros discussed in the main text. The position-dependent phases in Eqs.~\eqref{eq:AR2fSM} and \eqref{eq:AT2fSM} retain the full response of the finite array and no transverse-momentum conservation has been assumed.

\begin{figure}
\includegraphics[width=8.5cm]{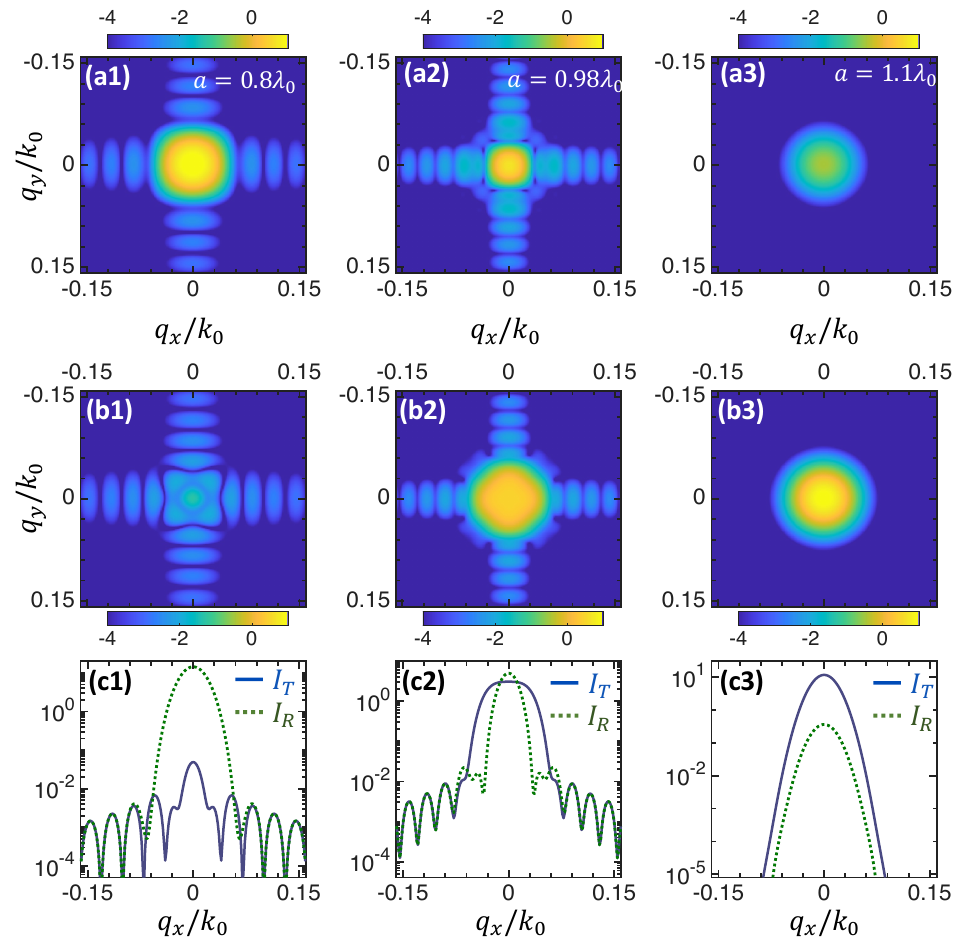}
\centering
\caption{Single-photon intensity distributions in the 2$f$ detection plane. The top and middle rows show the reflected and transmitted intensities, respectively, $\log_{10}I_{\alpha}(\boldsymbol{\rho}$, with ($\alpha=R,T$). Under the 2$f$ mapping, the detector coordinate $\boldsymbol{\rho}$ resolves the transverse momentum $\boldsymbol{q}=k_0\boldsymbol{\rho}/f$. The bottom row shows the corresponding central cuts $I_{\alpha}(q_x,q_y=0)$ on a logarithmic scale for reflected and transmitted photons. The three columns correspond to lattice constants $a=0.8\lambda_0$, $0.98\lambda_0$, and $1.1\lambda_0$, respectively. The other parameters are $L=40$ and $w_0=10\lambda_0$, with the collection aperture restricted to $|\boldsymbol{q}|<0.16k_0$.\label{fig:intensity}}
\end{figure}

Figure~\ref{fig:intensity} illustrates how the Rayleigh threshold reorganizes the single-photon momentum distribution. Below threshold, at $a=0.8\lambda_0$, strong collective reflection suppresses transmission through destructive interference. As $a$ approaches the threshold, the reflected intensity decreases while transmission recovers. Above threshold, at $a=1.1\lambda_0$, near-axis reflection is strongly suppressed, and transmission is dominated by a smooth central lobe. The interference fringes observed below and near threshold also disappear as the opening of additional diffraction channels broadens the collective modes and suppresses coherent scattering into the collection cone.

The fringes originate from interference among fields radiated by different atomic sites. For $L$ uniformly weighted lattice sites along either transverse direction, the corresponding structure factor is
\begin{equation}
\sum_{n=0}^{L-1}e^{-inq_\mu a}
=
e^{-i(L-1)q_\mu a/2}
\frac{\sin(Lq_\mu a/2)}
{\sin(q_\mu a/2)},
\qquad \mu=x,y.
\end{equation}
This gives a characteristic fringe spacing $\Delta q_\mu/k_0\simeq\lambda_0/(La)$. For $L=40$, the estimated spacings are $0.031$ and $0.026$ at $a=0.8\lambda_0$ and $0.96\lambda_0$, respectively. The Gaussian illumination and nonuniform collective response modify the fringe amplitudes, while their characteristic spacing remains governed by the finite spatial extent of coherent scattering.

An additional feature appears in transmission at $a=0.8\lambda_0$. A nearly zero-intensity ring absent in reflection [Fig.~\ref{fig:intensity}(b1)]. As shown by Eq.~(S100), this ring results from momentum-space Fano interference between the directly transmitted Gaussian field $\xi(\boldsymbol q)$ and the collectively scattered field. The corresponding intensity minima are resolved in Fig.~\ref{fig:intensity}(c1). Near threshold, the ring becomes distorted as the collective response is reorganized, whereas above threshold the weakened scattered field can no longer produce pronounced destructive interference with the directly transmitted field.

\subsection{Two-Photon scattering in the momentum-resolved 2$f$ configuration}

We now obtain the equal-time, transverse-momentum-resolved two-photon amplitude by projecting the plane-wave scattering matrix onto the $2f$
detection modes. Starting from the definition of the two-photon detection amplitude,
\begin{equation}
\mathcal A^{(2)}_{\alpha\alpha}(\boldsymbol\rho,\boldsymbol\rho')
=\langle 0|\hat E^{(+)}_\alpha(\boldsymbol\rho')
\hat E^{(+)}_\alpha(\boldsymbol\rho)\hat S|2_{\rm in}\rangle,
\label{eq:A2detect2fSM}
\end{equation}
we insert the $2f$ field operator of Eq.~\eqref{eq:E2fSM} and the normalized incident two-photon state defined in the preceding section. Explicitly, before taking the narrowband limit,
\begin{align}
\mathcal A^{(2)}_{\alpha\alpha}(\boldsymbol\rho,\boldsymbol\rho')
=&\frac{\tilde C_\alpha^{\,2}}{\sqrt2}
P(\boldsymbol q)P(\boldsymbol q')
\int_0^\infty d\omega_3\int_0^\infty d\omega_4\,
 e^{i[\phi_\alpha(k_{z3})+\phi_\alpha(k_{z4})]}\int d^2q_1\int d^2q_2
\int_0^\infty\!\frac{d\omega_1\,d\omega_2}
{\sqrt{v_{z1}v_{z2}}}
\nonumber\\
&\times \xi(\boldsymbol q_1)\xi(\boldsymbol q_2)
\chi(\omega_1)\chi(\omega_2)
\mathcal A^{(2)}(\boldsymbol k_3,\lambda_\alpha,
\boldsymbol k_4,\lambda_\alpha;
\boldsymbol k_1,-,\boldsymbol k_2,-),
\label{eq:A2projection2fSM}
\end{align}
where $\boldsymbol k_3=(\boldsymbol q,s_\alpha k_{z3})$ and
$\boldsymbol k_4=(\boldsymbol q',s_\alpha k_{z4})$, with
$\boldsymbol q=k_0\boldsymbol\rho/f$ and
$\boldsymbol q'=k_0\boldsymbol\rho'/f$.
The disconnected and connected parts of the plane-wave $S$ matrix therefore give
\begin{equation}
\mathcal A^{(2)}_{\alpha\alpha}(\boldsymbol\rho,\boldsymbol\rho')
=\mathcal A^{(2)}_{\alpha\alpha,{\rm dis}}(\boldsymbol\rho,\boldsymbol\rho')
+\mathcal A^{(2)}_{\alpha\alpha,{\rm con}}(\boldsymbol\rho,\boldsymbol\rho').
\end{equation}

\subsubsection{Disconnected contribution}
The plane-wave disconnected matrix element is the sum of the two products of one-photon $S$ matrices. Inserting this result into
Eq.~\eqref{eq:A2projection2fSM}, each product independently yields the single-photon detection amplitude derived in the previous subsection, including the direct transmission term when $\alpha=T$. Because the two incoming photons occupy the same wave-packet mode, the two permutations give identical
contributions. Thus
\begin{align}
\mathcal A^{(2)}_{\alpha\alpha,{\rm dis}}(\boldsymbol\rho,\boldsymbol\rho')
&=\frac{1}{\sqrt2}\left[
\mathcal A^{(1)}_\alpha(\boldsymbol\rho)
\mathcal A^{(1)}_\alpha(\boldsymbol\rho')
+\mathcal A^{(1)}_\alpha(\boldsymbol\rho')
\mathcal A^{(1)}_\alpha(\boldsymbol\rho)\right]=\sqrt2\,\mathcal A^{(1)}_\alpha(\boldsymbol\rho)
\mathcal A^{(1)}_\alpha(\boldsymbol\rho').
\label{eq:A2dis2fSM}
\end{align}
The factor $1/\sqrt2$ comes from the definition of the Fock state $|2_{\rm in}\rangle$. This is consistent with the weak coherent-state expansion in the preceding section. In particular, the amplitudes $\mathcal A^{(1)}_R$ and $\mathcal A^{(1)}_T$ in this equation are
exactly those of Eqs.~\eqref{eq:AR2fSM} and \eqref{eq:AT2fSM}.

\subsubsection{Connected contribution}
We next substitute the complete plane-wave connected amplitude of Eq.~\eqref{eq:A2conSM}, whose kernel is
$\mathcal M-\mathcal N$, with $\mathcal N=\mathcal N^{(13|24)}+\mathcal N^{(23|14)}$. The two incident transverse-momentum integrations yield $2\pi\psi_{j}$, while the outgoing plane-wave factors give $\exp[-i(\boldsymbol q\cdot\boldsymbol R_{j_3} +\boldsymbol q'\cdot\boldsymbol R_{j_4})]$. The energy-conserving delta function $\delta(\omega_3+\omega_4-\omega_1-\omega_2)$ fixes $\omega_4=\omega_1+\omega_2-\omega_3$. As in the one-photon calculation, the near-resonant coupling and continuum factors combine into $[3\gamma_0/(4k_0^2)]^2$. The two output helicity-overlap factors have the same combined sign for $RR$ and $TT$. We obtain
\begin{align}
\mathcal A^{(2)}_{\alpha\alpha,{\rm con}}(\boldsymbol\rho,\boldsymbol\rho')
= &-\frac{i}{\sqrt2}C_\alpha^2
\left(\frac{3\gamma_0}{4k_0^2}\right)^2
P(\boldsymbol q)P(\boldsymbol q')
\sum_{j_1j_2j_3j_4}
 e^{-i(\boldsymbol q\cdot\boldsymbol R_{j_3}
       +\boldsymbol q'\cdot\boldsymbol R_{j_4})}
\left[\mathcal M^{\rm eff}_{j_3j_4;j_1j_2}
-\mathcal N^{\rm eff}_{j_3j_4;j_1j_2}\right]
\psi_{j_1}\psi_{j_2} \\
& \equiv A^{(2)}_{\alpha\alpha,M} + A^{(2)}_{\alpha\alpha,N},
\label{eq:A2con2fSM}
\end{align}
Here $A^{(2)}_{\alpha\alpha,M}$ denotes two-excitation contribution, $A^{(2)}_{\alpha\alpha,N}$ arises from the principal-value term, and $C_\alpha$ is the same effective single-photon detection constant as in Eq.~\eqref{eq:Calpha}. Accordingly, the two effective kernels are normalized by the square
of this spectral integral:
\begin{align}
\mathcal M^{\rm eff}_{j_3j_4;j_1j_2}
={}&\frac{1}{\left[\int_0^\infty d\omega\,\chi(\omega)\right]^2}
\int_0^\infty d\omega_1\,\chi(\omega_1)
\int_0^\infty d\omega_2\,\chi(\omega_2)
\int_{-\infty}^{\infty}\frac{d\omega_3}{2\pi}\,
\mathcal M_{j_3j_4;j_1j_2}
(\omega_3,\omega_1+\omega_2-\omega_3;\omega_1,\omega_2),
\label{eq:MeffSM}\\
\mathcal N^{\rm eff}_{j_3j_4;j_1j_2}
={}&\frac{1}{\left[\int_0^\infty d\omega\,\chi(\omega)\right]^2}
\int_0^\infty d\omega_1\,\chi(\omega_1)
\int_0^\infty d\omega_2\,\chi(\omega_2)
\int_{-\infty}^{\infty}\frac{d\omega_3}{2\pi}\,
\mathcal N_{j_3j_4;j_1j_2}
(\omega_3,\omega_1+\omega_2-\omega_3;\omega_1,\omega_2).
\label{eq:NeffSM}
\end{align}
The output-frequency integrals here are coherent integrations of scattering amplitudes, appropriate to the ideal broadband, equal-time measurement. 
Within the near-resonant Markov approximation, the frequency integral over $\omega_3$ has been extended to the entire real axis because the scattering response is concentrated near the positive optical resonance frequency $\omega_0$, which is much larger than the relevant linewidths and collective frequency shifts, making contributions from the unphysical negative-frequency region negligible.

We first evaluate the principal-value contribution. The single-excitation Green matrix $\mathcal G(\omega) = 1/(\omega - \hat{H}^{(1)}_{\rm eff})$
defined in the single-photon scattering subsection, obeys the resolvent identity 
\begin{equation}
\mathcal G(\omega)-\mathcal G(\omega')
=(\omega'-\omega)\mathcal G(\omega)\mathcal G(\omega').
\label{eq:2fResolventIdentitySM}
\end{equation}
Using energy conservation, $\omega_1+\omega_2=\omega_3+\omega_4$, this identity removes the apparent singular denominators in Eqs.~\eqref{eq:A2K13SM} and \eqref{eq:A2K23SM}:
\begin{align}
\mathcal N^{(13|24)}_{j_3j_4;j_1j_2}
={}&[\mathcal G(\omega_4)\mathcal G(\omega_2)]_{j_4j_2}\mathcal
G_{j_3j_1}(\omega_1)
+\mathcal G_{j_4j_2}(\omega_2)
[\mathcal G(\omega_1)\mathcal G(\omega_3)]_{j_3j_1},
\nonumber\\
\mathcal N^{(23|14)}_{j_3j_4;j_1j_2}
= & [\mathcal G(\omega_3)\mathcal G(\omega_2)]_{j_3j_2}
\mathcal G_{j_4j_1}(\omega_1)
+\mathcal G_{j_3j_2}(\omega_2)
[\mathcal G(\omega_1)\mathcal G(\omega_4)]_{j_4j_1}.
\label{eq:2fNregularSM}
\end{align}
For the Markov resolvent, the symmetric equal-time frequency integrals are
\begin{equation}
\int_{-\infty}^{\infty}\frac{d\omega_3}{2\pi}
\mathcal G(\omega_3)=-\frac{i}{2} \mathcal I_N,
\qquad
\int_{-\infty}^{\infty}\frac{d\omega_3}{2\pi}
\mathcal G(\omega_4)=\int_{-\infty}^{\infty}\frac{d\omega_3}{2\pi}
\mathcal G(\omega_1+\omega_2-\omega_3)
=-\frac{i}{2}\mathcal I_N,
\label{eq:2fHalfPoleSM}
\end{equation}
with the $N\times N$ identity matrix $\mathcal I_N$. These identities follow by integrating each pole
$1/(\omega-E_n)$ with ${\rm Im}E_n<0$. Substitution into Eq.~\eqref{eq:2fNregularSM} gives
\begin{align}
\int_{-\infty}^{\infty}\frac{d\omega_3}{2\pi}
\mathcal N^{(13|24)}_{j_3j_4;j_1j_2}
&=-iG_{j_3j_1}(\omega_1)G_{j_4j_2}(\omega_2),
\nonumber\\
\int_{-\infty}^{\infty}\frac{d\omega_3}{2\pi}
\mathcal N^{(23|14)}_{j_3j_4;j_1j_2}
&=-iG_{j_3j_2}(\omega_2)G_{j_4j_1}(\omega_1).
\label{eq:2fNfrequencySM}
\end{align}
Thus, without any remaining outgoing-frequency integral,
\begin{align}
\mathcal N^{\rm eff}_{j_3j_4;j_1j_2}
= &-\frac{i}{\left[\int_0^\infty d\omega\,\chi(\omega)\right]^2}
\int_0^\infty d\omega_1\,\chi(\omega_1)
\int_0^\infty d\omega_2\,\chi(\omega_2)\left[\mathcal G_{j_3j_1}(\omega_1)G_{j_4j_2}(\omega_2)
+\mathcal G_{j_3j_2}(\omega_2)G_{j_4j_1}(\omega_1)\right]
\nonumber\\
\simeq &-i\left[\mathcal G_{j_3j_1}(\omega_{\rm in})\mathcal G_{j_4j_2}(\omega_{\rm in})
+\mathcal G_{j_3j_2}(\omega_{\rm in})\mathcal G_{j_4j_1}(\omega_{\rm in})\right].
\label{eq:NeffAnalyticSM}
\end{align}
The second line uses a narrowband incident pulse. We note that $\mathcal N^{\rm eff}$ is the finite remainder of
the alternating absorption--emission orderings after the separately energy-conserving terms have been assigned to
$\mathcal A^{(2)}_{\rm dis}$.

For the four orderings that propagate through the physical two-excitation subspace, the two outgoing permutations of $\mathcal M$ are equal after frequency integration because $[\hat\sigma_{j_3},\hat\sigma_{j_4}]=0$. Equation~\eqref{eq:2fHalfPoleSM} then gives 
\begin{align}
\int_{-\infty}^{\infty}\frac{d\omega_3}{2\pi}
\mathcal M_{j_3j_4;j_1j_2}
(\omega_3,\omega_1+\omega_2-\omega_3;\omega_1,\omega_2) =-i\langle G|\hat\sigma_{j_4}\hat\sigma_{j_3}
\frac{1}{E-\hat H_{\rm eff}^{(2)}}
\left[
\hat\sigma_{j_2}^\dagger
\frac{1}{\omega_1-\hat H_{\rm eff}^{(1)}}\hat\sigma_{j_1}^\dagger
+\hat\sigma_{j_1}^\dagger
\frac{1}{\omega_2-\hat H_{\rm eff}^{(1)}}\hat\sigma_{j_2}^\dagger
\right]|G\rangle,
\label{eq:2fMfrequencySM}
\end{align}
Hence
\begin{align}
\mathcal M^{\rm eff}_{j_3j_4;j_1j_2}
&=-\frac{i}{\left[\int_0^\infty d\omega\,\chi(\omega)\right]^2}
\int_0^\infty d\omega_1\,\chi(\omega_1)
\int_0^\infty d\omega_2\,\chi(\omega_2)
\langle G|\hat\sigma_{j_4}\hat\sigma_{j_3}
\frac{1}{\omega_1+\omega_2-\hat H_{\rm eff}^{(2)}}
\left[
\hat\sigma_{j_2}^\dagger
\frac{1}{\omega_1-\hat H_{\rm eff}^{(1)}}\hat\sigma_{j_1}^\dagger
+\hat\sigma_{j_1}^\dagger
\frac{1}{\omega_2-\hat H_{\rm eff}^{(1)}}\hat\sigma_{j_2}^\dagger
\right]|G\rangle
\nonumber\\
\simeq{}&-i\langle G|\hat\sigma_{j_4}\hat\sigma_{j_3}
\frac{1}{2\omega_{\rm in}-\hat H_{\rm eff}^{(2)}}
\left[
\hat\sigma_{j_2}^\dagger
\frac{1}{\omega_{\rm in}-\hat H_{\rm eff}^{(1)}}\hat\sigma_{j_1}^\dagger
+\hat\sigma_{j_1}^\dagger
\frac{1}{\omega_{\rm in}-\hat H_{\rm eff}^{(1)}}\hat\sigma_{j_2}^\dagger
\right]|G\rangle.
\label{eq:MeffAnalyticSM}
\end{align}
The outgoing-frequency identities above are exact within the frequency-independent Markov model. The final narrowband approximations apply only to the incident spectral integrals.

\subsubsection{Single-atom consistency check and exact cancellation of the disconnected two-photon amplitude}
As a benchmark, consider a single ideal two-level atom with $ \mathcal G(\omega)=1/(\omega-h)$ and $h=\omega_0-i\gamma_0/2$. The physical two-excitation subspace is empty, so $\mathcal M^{\rm eff} =0$. Nevertheless, the alternating absorption--emission orderings give a nonzero frequency-resolved kernel. Inserting the scalar Green function into Eq.~\eqref{eq:NeffAnalyticSM}, we obtain
\begin{equation}
 \mathcal N^{\rm eff}=-\frac{2i}{(\Delta+i\gamma_0/2)^2}.
 \label{eq:2fSingleAtomEffectiveSM}
\end{equation}
The reflected one-photon amplitude in Eq.~\eqref{eq:AR2fSM} then reduces to
\begin{equation}
 \mathcal A_R^{(1)}(\boldsymbol\rho)
 =iC_R\frac{3\gamma_0}{4k_0^2}P(\boldsymbol q)
 \frac{e^{-i\boldsymbol q\cdot\boldsymbol R_1}\psi_1}{\Delta+i\gamma_0/2}.
 \label{eq:2fSingleAtomOnePhotonSM}
\end{equation}
Consequently, the disconnected contribution is
\begin{align}
 \mathcal A_{RR,{\rm dis}}^{(2)}(\boldsymbol\rho,\boldsymbol\rho')
 &=-\sqrt2\,C_R^2\left(\frac{3\gamma_0}{4k_0^2}\right)^2
 P(\boldsymbol q)P(\boldsymbol q')
 \frac{e^{-i(\boldsymbol q+\boldsymbol q')\cdot\boldsymbol R_1}\psi_1^2}{(\Delta+i\gamma_0/2)^2},
 \label{eq:2fSingleAtomDisconnectedSM}\\
 \mathcal A_{RR,{\rm con}}^{(2)}(\boldsymbol\rho,\boldsymbol\rho')
 &=\mathcal A_{RR,N}^{(2)}=+\sqrt2\,C_R^2\left(\frac{3\gamma_0}{4k_0^2}\right)^2
 P(\boldsymbol q)P(\boldsymbol q')
 \frac{e^{-i(\boldsymbol q+\boldsymbol q')\cdot\boldsymbol R_1}\psi_1^2}{(\Delta+i\gamma_0/2)^2}.
 \label{eq:2fSingleAtomConnectedSM}
\end{align}
Here the connected amplitude follows from Eq.~\eqref{eq:A2con2fSM} with the full kernel $\mathcal M^{\rm eff}-\mathcal N^{\rm eff}$.
The disconnected and connected contributions therefore cancel exactly,
\begin{equation}
 \mathcal A_{RR}^{(2)}(\boldsymbol\rho,\boldsymbol\rho')=0,
 \qquad
 g_{RR}^{(2)}(\boldsymbol\rho,\boldsymbol\rho';0)=0.
 \label{eq:2fSingleAtomCancellationSM}
\end{equation}
This result does not imply that the frequency-resolved plane-wave connected amplitude vanishes. Indeed, Eq.~\eqref{eq:2fSingleAtomConnectedSM} remains nonzero even though the atomic two-excitation subspace is absent.

Importantly, this cancellation is not restricted to the single-atom limit. For an arbitrary atomic array, the disconnected reflected two-photon amplitude
$\mathcal A_{RR,{\rm dis}}^{(2)}$ continues to cancel the principal-value contribution $\mathcal A_{RR,N}^{(2)}$ to the connected amplitude. Within the Markov and narrowband limits adopted above, we obtain
\begin{align}
\mathcal A^{(2)}_{RR,N}(\boldsymbol\rho,\boldsymbol\rho')
={}&\frac{iC_R^2}{\sqrt{2}}
\left(\frac{3\gamma_0}{4k_0^2}\right)^2
P(\boldsymbol q)P(\boldsymbol q')
\sum_{j_1j_2j_3j_4}
e^{-i(\boldsymbol q\cdot\boldsymbol R_{j_3}
+\boldsymbol q'\cdot\boldsymbol R_{j_4})}
N^{\rm eff}_{j_3j_4;j_1j_2}
\psi_{j_1}\psi_{j_2}
\nonumber\\
={}&-\sqrt{2}\,
\mathcal A_R^{(1)}(\boldsymbol\rho)
\mathcal A_R^{(1)}(\boldsymbol\rho')
= -\mathcal A^{(2)}_{RR,{\rm dis}}
(\boldsymbol\rho,\boldsymbol\rho').
\label{eq:NdisCancellationSM}
\end{align}
The cancellation therefore holds at the amplitude level for arbitrary detected transverse momenta within the collection aperture, rather than only after integration over momentum. This is consistent with photon scattering in a one-dimensional waveguide as shown in Eq.~(D3) in Ref.~\cite{yang2024quantum}. 

This exact cancellation relies on equal-time, frequency-integrated detection together with the frequency-independent Markov propagator and the narrowband and low-NA approximations used above. It does not generally persist for frequency-resolved or unequal-time correlations. In transmission, the direct incident-field contribution to $\mathcal A_T^{(1)}$ prevents an analogous cancellation of the full disconnected two-photon amplitude.

\subsection{Momentum-resolved two-photon correlations and nonlinear contrast}
Combining Eqs.~\eqref{eq:A2dis2fSM} and \eqref{eq:A2con2fSM}, the normalized, transverse-momentum-resolved equal-time correlation function becomes
\begin{equation}
 g^{(2)}_{\alpha\alpha}(\boldsymbol\rho,\boldsymbol\rho')
\equiv  \frac{\left|\mathcal{A}^{(2)}_{\alpha\alpha}(\boldsymbol{\rho},\boldsymbol{\rho}')/\sqrt{2}\right|^2}{|\mathcal{A}_\alpha^{(1)}(\boldsymbol{\rho})|^2|\mathcal{A}_\alpha^{(1)}(\boldsymbol{\rho}')|^2}=\left|1+
\frac{\mathcal A^{(2)}_{\alpha\alpha,{\rm con}}(\boldsymbol\rho,\boldsymbol\rho')}
{\sqrt2\,\mathcal A^{(1)}_\alpha(\boldsymbol\rho)
\mathcal A^{(1)}_\alpha(\boldsymbol\rho')}\right|^2,
\label{eq:g2finalSM}
\end{equation}
where the factor $\sqrt{2}$ follows from the two-photon component of the weak coherent-state expansion in Eq.~\eqref{eq:coherentstate}. The complete connected amplitude contains contributions from both the two-excitation propagation and the alternating absorption--emission orderings. The principal-value correction must therefore be retained in calculations of $g^{(2)}$ and the integrated nonlinear contrasts.

\begin{figure}
\includegraphics[width=8.5cm]{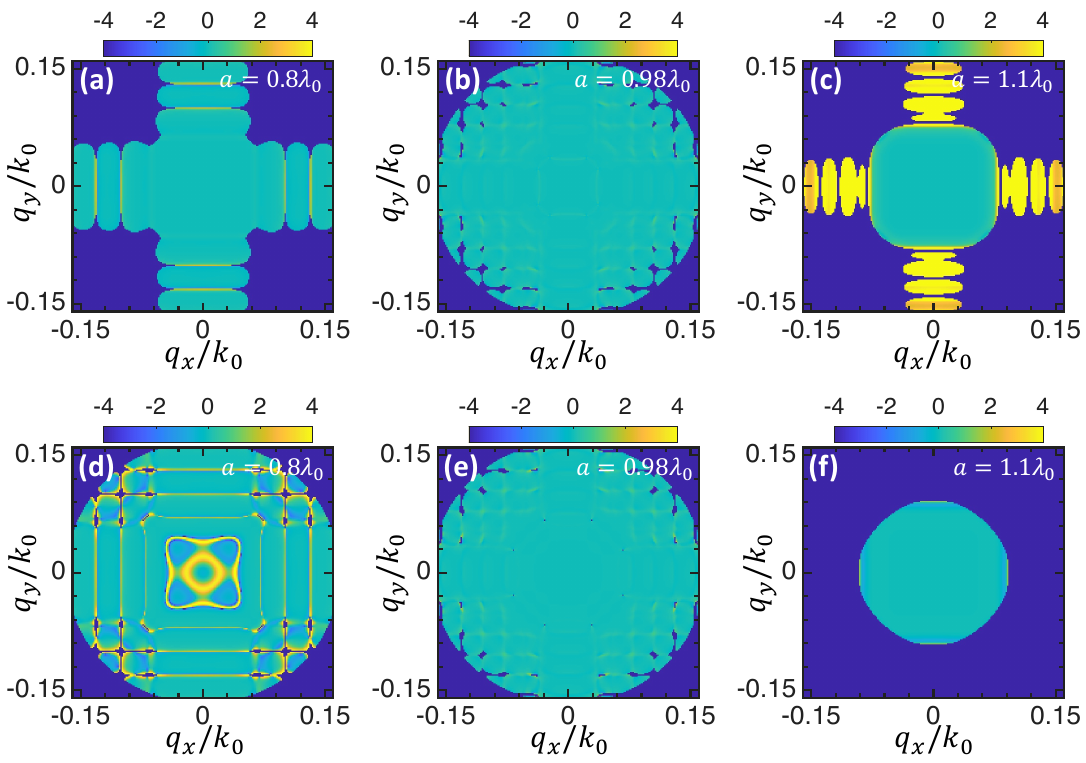}
\centering
\caption{Momentum-resolved second-order photon correlations in the 2$f$ detection plane. The top and bottom rows show $\log_{10}g_{\alpha\alpha}^{(2)}(\boldsymbol q,\boldsymbol q)$ for reflected $\alpha=R$ and transmitted $\alpha=T$ photons, respectively, with both photons detected in the same transverse momentum mode.  The three columns correspond to lattice constants $a=0.8\lambda_0$, $0.98\lambda_0$, and $1.1\lambda_0$, respectively. The other parameters are $L=40$ and $w_0=10\lambda_0$, with the collection aperture restricted to $|\boldsymbol q|<0.16k_0$. \label{fig:FigS_g2_sameq}}
\end{figure}

\subsubsection{Equal-momentum photon correlations}
Figure~\ref{fig:FigS_g2_sameq} shows the equal-momentum photon correlations $g_{\alpha\alpha}^{(2)}(\boldsymbol q,\boldsymbol q)$ across the Rayleigh threshold. Below threshold ($a=0.8\lambda_0$), both reflected and transmitted correlations exhibit pronounced square-symmetric interference fringes. These structures are strongly suppressed near the threshold ($a=0.98\lambda_0$), where $g_{\alpha\alpha}^{(2)}$ approaches unity over much of the collection aperture. Above threshold ($a=1.1\lambda_0$), stripe-like features
reappear in reflection, while the transmitted correlations become nearly uniform. The characteristic fringe spacing can be understood from the same finite-lattice interference factor in single-photon intensities. 

The transmitted correlations exhibit an additional ring-like enhancement below threshold [Fig.~\ref{fig:FigS_g2_sameq}(d)], originating from the Fano minima in the single-photon transmission intensity [Fig.~\ref{fig:intensity}(b1)]. As shown in Eq.~(S100), destructive interference between the directly transmitted and collectively scattered fields
produces near-zero transmission amplitudes at finite transverse momenta. The resulting suppression of $I_T(\boldsymbol q)$ enhances the normalized correlation $g_{TT}^{(2)}(\boldsymbol q,\boldsymbol q) =G_{TT}^{(2)}(\boldsymbol q,\boldsymbol q)/ [2I_T^2(\boldsymbol q)]$, provided the coincidence signal does not vanish at the same rate. Near the Rayleigh threshold, the Fano minima are strongly modified and the associated ring-like structure disappears. Similarly, the large normalized reflected correlations above threshold occur where the single-photon reflected intensity is strongly suppressed, and therefore do not necessarily imply enhanced absolute two-photon scattering.

\subsubsection{Joint momentum correlations along the $x$ direction}

\begin{figure}
\includegraphics[width=8.5cm]{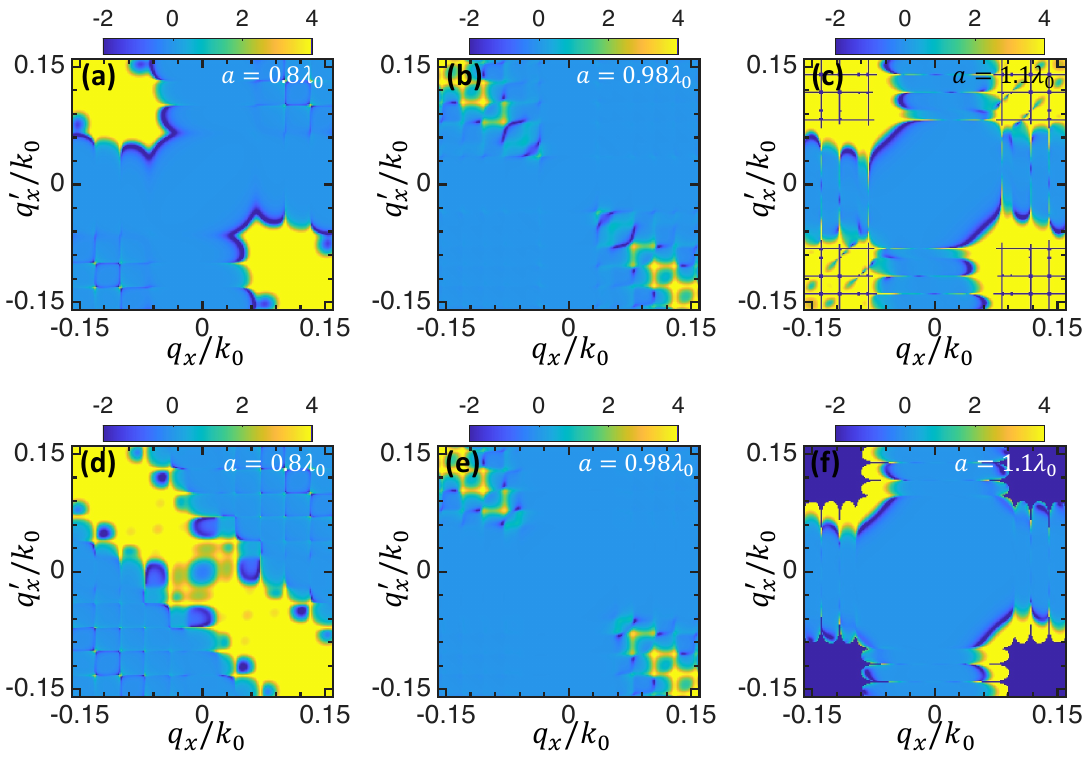}
\centering
\caption{Momentum correlations along the $x$ direction. The top and bottom rows show $\log_{10}g_{\alpha\alpha}^{(2)}(q_{x},0;q'_{x},0)$ for reflected  and transmitted  photons, respectively, with both photons detected along $q_{y}=q'_{y}=0$. The horizontal and vertical axes correspond to the transverse momenta $q_{x}$ and $q'_{x}$ of the two detected photons, respectively. The three columns correspond to lattice constants $a=0.8\lambda_0$, $0.98\lambda_0$, and $1.1\lambda_0$, respectively. The other parameters are $L=40$ and $w_0=10\lambda_0$, with the collection aperture restricted to $|\boldsymbol q|,|\boldsymbol q'|<0.16k_0$. \label{fig:FigS_g2_q1xq2x}}
\end{figure}

Figure~\ref{fig:FigS_g2_q1xq2x} shows the evolution of the equal-time momentum correlations $g_{\alpha\alpha}^{(2)}(q_{x},0;q'_{x},0)$ across the
Rayleigh threshold, resolving correlations between distinct transverse momentum modes. Below threshold $a=0.8\lambda_0$, both reflected and transmitted photons exhibit pronounced opposite-momentum correlations along $q_{x}+q'_{x}\simeq0$, accompanied by fine interference fringes. Near the threshold $a=0.98\lambda_0$, these structures are strongly reorganized and become more localized in momentum space. Above threshold $a=1.1\lambda_0$, enhanced correlations reappear in reflection, predominantly at off-axis momenta, whereas transmission develops a broad central region with $g^{(2)}\approx 1$ surrounded by enhanced correlations.

The opposite-momentum correlations originate from coherent scattering by collectively excited atomic pairs, which favors approximate conservation of their total transverse momentum. Finite array size and beam waist broaden this momentum-matching condition, while interference across the array produces fringes with a characteristic momentum scale of order $k_0\lambda_0/(La)$. Their detailed structure is governed by the collective scattering amplitudes. In particular, the disconnected reflected amplitude exactly cancels the $N$ contribution under the approximations discussed in the Supplemental Material, leaving the two-excitation $M$ term to determine the reflected coincidence amplitude.

In transmission, the additional narrow bands and localized enhancements below threshold [Fig.~\ref{fig:FigS_g2_q1xq2x}(d)] are associated with the
single-photon Fano minima in Fig.~\ref{fig:intensity}(b1). When either detected momentum approaches a transmission zero, the suppressed single-photon
intensity enhances the normalized $g_{TT}^{(2)}$, producing the joint-momentum counterparts of the features observed in the equal-momentum correlations. As the Rayleigh threshold is crossed, the collective response and Fano minima are restructured, substantially modifying these correlation patterns.

\section{Validity of the two-level reduction and Markov approximation} 

\subsection{Validity of the two-level approximation for specular reflection}
\label{sec:multilevel_check_SM}

To examine the effective two-level approximation (TLA), we consider the $^{1}S_0\!\rightarrow{}^{3}P_1$ transition of $^{88}\mathrm{Sr}$, retaining
all three excited Zeeman sublevels $|e_{j,m}\rangle$ ($m=0,\pm1$) of each atom. A static magnetic field $\boldsymbol B=B\boldsymbol e_z$
is applied normal to the array. Defining $\hat\sigma_{j,m}=|g_j\rangle\langle e_{j,m}|$ and $\delta_Z=g_J\mu_B B/\hbar$, the atomic Hamiltonian is
\begin{equation}
\hat H_{\mathrm{atom}}^{(J=1)}
=\hbar\sum_{j=1}^{N}\sum_{m=-1}^{1}
(\omega_0+m\delta_Z)\hat\sigma_{j,m}^{\dagger}\hat\sigma_{j,m},
\label{eq:Hatom_multilevel_SM}
\end{equation}
where $g_J\simeq3/2$ and $\omega_0$ is the zero-field transition frequency. The full electric-dipole coupling to the free-space field is
\begin{equation}
\hat H_{\mathrm{int}}^{(J=1)}
=-\sum_j\hat{\boldsymbol d}_j\cdot
\hat{\boldsymbol E}(\boldsymbol R_j),\qquad
\hat{\boldsymbol d}_j
=\sum_{m=-1}^{1}
\bigl(\boldsymbol d_m^*\hat\sigma_{j,m}^{\dagger}
+\boldsymbol d_m\hat\sigma_{j,m}\bigr),
\label{eq:Hint_multilevel_SM}
\end{equation}
with $\boldsymbol d_m=d\boldsymbol e_m$,
$\boldsymbol e_{\pm}=(\boldsymbol e_x\pm i\boldsymbol e_y)/\sqrt{2}$,
and $\boldsymbol e_0=\boldsymbol e_z$.

Eliminating the electromagnetic field within the Born--Markov approximation yields the effective non-Hermitian Hamiltonian in scattering processes,
\begin{align}
\hat H_{\mathrm{eff}}^{(J=1)}
={}&\sum_{j,m}
\left(\omega_0+m\delta_Z-\frac{i\gamma_0}{2}\right)
\hat{\sigma}^{\dagger}_{j,m}\hat{\sigma}_{j,m}+\sum_{i\ne j}\sum_{m,n=-1}^{1}
\left(J_{im,jn}-\frac{i\gamma_{im,jn}}{2}\right)\hat{\sigma}^{\dagger}_{i,m}\hat{\sigma}_{j,n},
\label{eq:Heff_multilevel_SM}
\end{align}
where 
\begin{align}
J_{im,jn}
&=-\frac{\omega_0^2}{\hbar\epsilon_0c^2}
\boldsymbol d_m^*\cdot
\operatorname{Re}\overleftrightarrow G(\boldsymbol R_i,\boldsymbol R_j;\omega_0)
\cdot\boldsymbol d_n,\nonumber\\
\gamma_{im,jn}
&=\frac{2\omega_0^2}{\hbar\epsilon_0c^2}
\boldsymbol d_m^*\cdot
\operatorname{Im}\overleftrightarrow G(\boldsymbol R_i,\boldsymbol R_j;\omega_0)
\cdot\boldsymbol d_n.
\label{eq:Jgamma_multilevel_SM}
\end{align}
The on-site radiative coupling is diagonal, $\gamma_{im,in}=\gamma_0\delta_{mn}$. We retain the off-diagonal $m\ne n$ collective couplings in Eq.~\eqref{eq:Heff_multilevel_SM} when evaluating the full multilevel response. The use of the same Green tensor at $\omega_0$ for all three transitions is justified by $\delta_Z\ll\omega_0$ over the optical bandwidth of interest.

The plane-wave single-photon scattering formula has the same input--output structure as Eq.~\eqref{eq:A1SM}, but its extension requires both the multilevel resolvent and the corresponding polarization-dependent coupling vertices:
\begin{align}
&\mathcal A^{(1)}(\boldsymbol p,\lambda';\boldsymbol k,\lambda)
=\delta_{\lambda\lambda'}\delta^{(3)}(\boldsymbol p-\boldsymbol k)
-2\pi i\,\delta(\omega_{\boldsymbol p}-\omega_{\boldsymbol k})
\mathcal T^{(J=1)}(\boldsymbol p,\lambda';\boldsymbol k,\lambda),
\nonumber\\[2pt]
&\mathcal T^{(J=1)}(\boldsymbol p,\lambda';\boldsymbol k,\lambda)
=\sum_{i,j}\sum_{m,n=-1}^{1}
\bigl(g_{\boldsymbol p,\lambda'}^{(m)}\bigr)^*
 g_{\boldsymbol k,\lambda}^{(n)}
 e^{-i(\boldsymbol p\cdot\boldsymbol R_i
       -\boldsymbol k\cdot\boldsymbol R_j)}
\left\langle G\right|\hat\sigma_{i,m}
\frac{1}{\omega_{\boldsymbol k}-\hat H_{\mathrm{eff}}^{(J=1)}}
\hat\sigma_{j,n}^{\dagger}\left|G\right\rangle,
\label{eq:T_multilevel_SM}
\end{align}
where
$g_{\boldsymbol k,\lambda}^{(m)}
=-i\sqrt{\omega_{\boldsymbol k}/[2\hbar\epsilon_0(2\pi)^3]}
\,\boldsymbol e_{\boldsymbol k,\lambda}^*\cdot\boldsymbol d_m$, consistent with the convention used for the two-level model. The Gaussian input and specular output modes are projected in exactly the same manner as in the two-level calculation.

The close agreement of the two models has a symmetry-based explanation. Since all atoms lie in the $xy$ plane, the free-space Green tensor has $G_{xz}=G_{yz}=0$ for every interatomic separation. Thus the $\pi$ transition is exactly decoupled from the in-plane $\sigma^\pm$ transitions by the collective couplings; its direct excitation also vanishes for an ideal normally incident transverse plane wave. For the in-plane transitions, the $\sigma^+\!\leftrightarrow\!\sigma^-$ coupling carries an azimuthal factor $e^{\pm2i\varphi_{ij}}$. The fourfold symmetry of the square array makes its matrix element between equally spatially symmetric $\sigma^+$ and $\sigma^-$ collective modes vanish. A finite-waist Gaussian beam contains small nonzero transverse momenta and can couple the driven $\sigma^+$ mode to spatially structured $\sigma^-$ modes. Likewise, a fully vectorial paraxial field has a weak longitudinal component. These effects are not strictly forbidden, but their contribution to the collected specular amplitude is small for the parameters studied here.

\begin{figure}
\includegraphics[width=8.5cm]{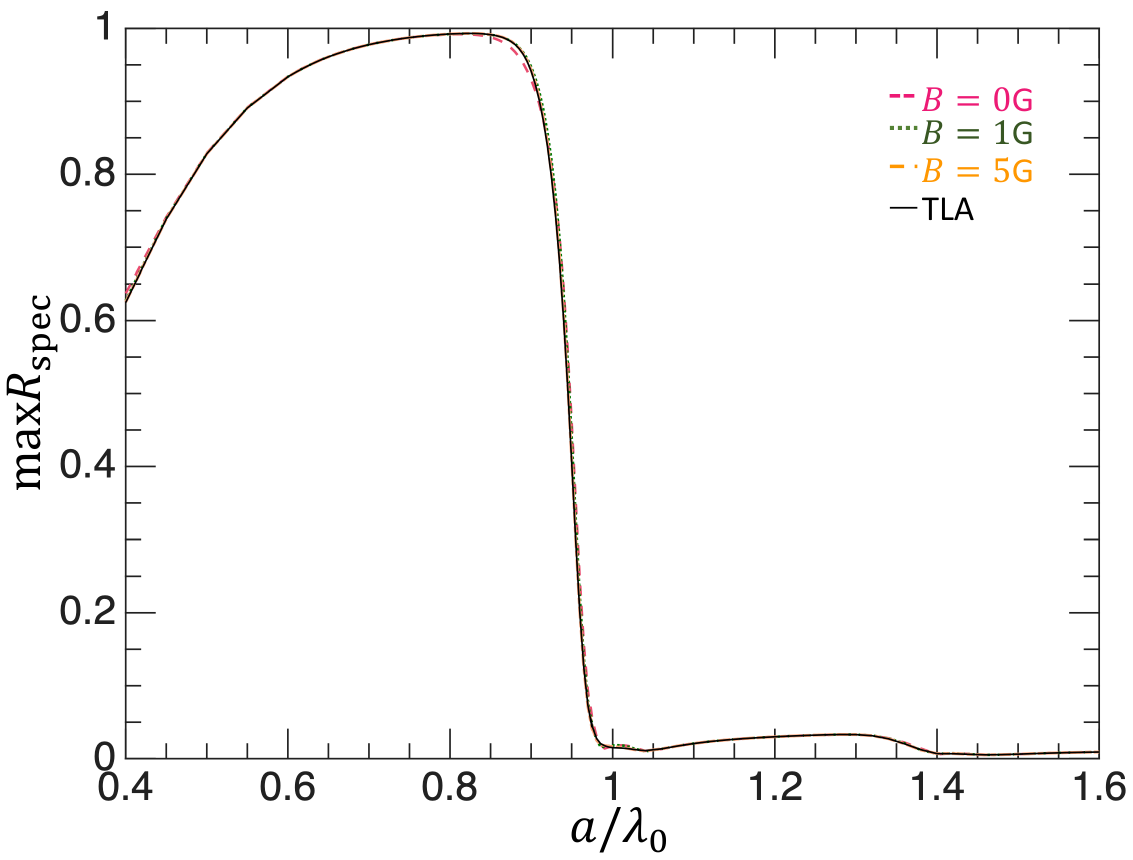}
\centering
\caption{Frequency-optimized specular reflectance of the full
$J=0\rightarrow1$ model for $B=0$, $1$, and $5\,\mathrm G$
compared with the effective two-level approximation (TLA).
The sharp loss of specular reflection near the Rayleigh
threshold $a=\lambda_0$ is insensitive to the additional
Zeeman sublevels for the parameters considered.}
\label{fig:multilevel_check_SM}
\end{figure}

Figure~\ref{fig:multilevel_check_SM} compares the frequency-optimized  specular reflectance $\max R_{\mathrm{spec}}$ of the full $J=0\!\rightarrow\!J=1$ model at $B=0$, $1$, and $5\,\mathrm{G}$ with that of the TLA. The incident frequency is optimized relative to the addressed $\sigma^+$ transition for each magnetic field and lattice spacing. All curves nearly coincide, reproducing both the high reflectance below threshold and its sharp collapse near $a=\lambda_0$; the zero-field curve shows only a small deviation near the transition. For $^{88}\mathrm{Sr}$, $\gamma_0/(2\pi)\simeq7.5\,\mathrm{kHz}$ and $\delta_Z/(2\pi)\simeq2.10\,\mathrm{MHz}\,(B/\mathrm G)$, so even $B=1\,\mathrm G$ separates adjacent Zeeman levels by about $280\gamma_0$, strongly suppressing off-resonant mixing. The agreement at $B=0$ reflects the symmetry and mode selectivity
of this particular single-photon specular observable, not an exact elimination of the other internal levels. The figure therefore validates the effective two-level calculation of the optimized linear reflectance, whereas a large Zeeman splitting
provides the controlled justification for the two-level reduction in nonlinear two-photon scattering.

\subsection{Exact one-photon plane-wave amplitude beyond Markov}
\label{sec:nonmarkSM}

For completeness, we now eliminate the three-dimensional radiation field without making the Markov approximation, following the single-excitation path-integral or Feshbach-resolvent procedure of Ref.~\cite{Shi2015multiphoton}.  This elimination is exact within the microscopic excitation-number-conserving Hamiltonian \eqref{eq:HintSM}. It does not require the quantum regression theorem or a time-local atomic master equation. The retarded Feshbach self-energy is
\begin{align}
\Sigma^R_{ij}(\omega)
&=\sum_{\lambda}\int d\boldsymbol k\,
\frac{|g_{\boldsymbol k,\lambda}|^2
 e^{i\boldsymbol q\cdot(\boldsymbol R_i-\boldsymbol R_j)}}
{\omega-\omega_{\boldsymbol q}+i0^+}=\mathcal P\int_0^\infty d\nu\,
 \frac{\mathcal J_{ij}(\nu)}{\omega-\nu}
-i\pi\mathcal J_{ij}(\omega)
\quad(\omega>0).
\label{eq:SigmaExactSM}
\end{align}
The diagonal real part (the Lamb shift) is renormalized into the measured atomic transition frequency. this is understood in the following equations. The exact RWA-model retarded single-excitation Green matrix is
\begin{equation}
\mathcal G^R(\omega)=\left[\omega\boldsymbol I-\omega_0\boldsymbol I-\Sigma^R(\omega)\right]^{-1}.
\label{eq:GRNM_SM}
\end{equation}
For reference, its time-domain retarded kernel is
\begin{align}
\Sigma^R_{ij}(t)
=-i\Theta(t)\sum_\lambda\int d\boldsymbol k\,
 |g_{\boldsymbol k,\lambda}|^2
 e^{i\boldsymbol k\cdot(\boldsymbol R_i-\boldsymbol R_j)}
 e^{-i\omega_{\boldsymbol k}t}.
\label{eq:memorySM}
\end{align}
This kernel is not proportional to $\delta(t)$ in general. The eliminated field mediates a time-delayed, frequency-dependent self-energy.

Within the RWA Hamiltonian, the exact non-Markov plane-wave single-photon $S$ matrix is
\begin{align}
\mathcal A^{(1)}_{\rm NM}(\boldsymbol p,\lambda';\boldsymbol k,\lambda)
&=\delta_{\lambda\lambda'}\delta^{(3)}(\boldsymbol p-\boldsymbol k)
-2\pi i\,\delta(\omega_{\boldsymbol p}-\omega_{\boldsymbol k})
\mathcal T_{\rm NM}(\boldsymbol p,\lambda';\boldsymbol k,\lambda),
\label{eq:ANMSM}\\
\mathcal T_{\rm NM}(\boldsymbol p,\lambda';\boldsymbol k,\lambda)
&=\sum_{ij}g^*_{\boldsymbol p,\lambda'}g_{\boldsymbol k,\lambda}
 e^{-i(\boldsymbol p\cdot\boldsymbol R_i-\boldsymbol k\cdot\boldsymbol R_j)}
 \mathcal G^R_{ij}(\omega_{\boldsymbol k}).
\label{eq:TNMSM}
\end{align}
Unlike the Markov result, both the propagator and the external couplings are evaluated at the actual photon frequency.

The common optical Green-tensor non-Markov approximation consists of replacing the renormalized off-site self-energy by the frequency-dependent retarded dyadic Green tensor,
\begin{equation}
\Sigma^R_{ij}(\omega)
\simeq-\frac{\omega^2}{\hbar\varepsilon_0c^2}
\boldsymbol d^*\cdot\overleftrightarrow G
(\boldsymbol R_i,\boldsymbol R_j;\omega)\cdot\boldsymbol d
\quad(i\ne j).
\label{eq:SigmaGoptSM}
\end{equation}
Eq.~\eqref{eq:SigmaGoptSM} is not an exact equality for the strictly positive-frequency RWA self-energy \eqref{eq:SigmaExactSM}, unless the missing virtual-frequency contribution is supplied.  This distinction is separate from the Markov approximation.  In the latter, $\Sigma^R(\omega)$ is evaluated at $\omega_0$, so that
\begin{equation}
\mathcal G^R(\omega)\ \longrightarrow\
\bigl[\omega-\hat H_{\rm eff}^{(1)}\bigr]^{-1},
\qquad
\Sigma^R(\omega_0)
\simeq \boldsymbol J-i\boldsymbol\gamma/2.
\label{eq:recoverMarkovSM}
\end{equation}
Thus the Markov limit of Eq.~\eqref{eq:TNMSM} recovers Eq.~\eqref{eq:T_SM} in the same optical Green-tensor convention.

\subsection{Markov approximation near the Rayleigh threshold}
\label{sec:MarkovcriterionSM}
We consider quasi-monochromatic paraxial pulses centered at
$\omega_{\rm in}=\omega_0+\Delta$, with an incident bandwidth
smaller than $\gamma_0$. Within the optical Green-tensor
prescription discussed above, the leading frequency dependence
of the atomic self-energy can be assessed by evaluating the
dyadic Green tensor at the incident frequency. Specifically,
for $i\ne j$, the off-diagonal coupling in the effective
Hamiltonian is modified according to
\begin{equation}
-\frac{\omega_0^2}{\hbar\epsilon_0c^2}
\boldsymbol d^*\!\cdot\!
\overleftrightarrow{\boldsymbol G}
(\boldsymbol R_i,\boldsymbol R_j;\omega_0)
\!\cdot\!\boldsymbol d
\;\longrightarrow\;
-\frac{\omega_{\rm in}^2}{\hbar\epsilon_0c^2}
\boldsymbol d^*\!\cdot\!
\overleftrightarrow{\boldsymbol G}
(\boldsymbol R_i,\boldsymbol R_j;\omega_{\rm in})
\!\cdot\!\boldsymbol d.
\label{eq:MarkovFrequencyCorrectionSM}
\end{equation}
This substitution captures the leading frequency dependence
within the optical Green-tensor approximation and should not
be confused with an exact non-Markovian elimination of the
microscopic RWA Hamiltonian.

In all calculations, the detuning is restricted to
$|\Delta|<2\gamma_0\ll\omega_0$. The frequency dependence
of the free-space dyadic Green tensor arises primarily from
the propagation phase
$e^{i\omega|\boldsymbol R_i-\boldsymbol R_j|/c}$,
together with smoothly varying prefactors. The Markov
approximation is therefore justified when
\begin{equation}
\frac{|\Delta|}{\omega_0}\ll1,\qquad
\frac{|\Delta|\,|\boldsymbol R_i-\boldsymbol R_j|}{c}
\ll1.
\label{eq:MarkovCriterionSM}
\end{equation}
For the Gaussian beam waist $w_0=10\lambda_0$,
the phase variation across one beam waist satisfies
\begin{equation}
\frac{|\Delta|w_0}{c}
<40\pi\frac{\gamma_0}{\omega_0}\ll1.
\label{eq:MarkovEstimateSM}
\end{equation}

The negligible propagation delay can also be verified
quantitatively from the collective dynamics. For an
$L\times L$ square array, the characteristic propagation
time across one side is $(L-1)a/c$. The relevant
dimensionless retardation parameter is
\begin{equation}
\frac{\Gamma_{\rm dom}(L-1)a}{c}
\lesssim 10^{-7}\ll1,
\label{eq:MarkovRetardationSM}
\end{equation}
where $\Gamma_{\rm dom}$ is the radiative decay rate of the
dominant collective mode. For the proposed $^{88}$Sr
implementation, $\lambda_0=689\,{\rm nm}$ and
$\gamma_0/2\pi\simeq7.5\,{\rm kHz}$. With $L=40$ and
$a\simeq\lambda_0$, even taking
$\Gamma_{\rm dom}=20\gamma_0$ gives
$\Gamma_{\rm dom}(L-1)a/c\simeq8.4\times10^{-8}$.
Thus, photon propagation across the array is much faster
than the collective atomic dynamics, quantitatively
justifying the neglect of propagation-induced memory effects.

Although the connected two-photon amplitude involves
integrations over outgoing frequencies, the relevant
resonant response is governed by collective linewidths
that are extremely small compared with the optical
frequency. Consequently, retaining the detuning in the
atomic resolvents while evaluating the dipole--dipole
couplings at $\omega_0$ provides an excellent approximation
for the present calculations.

Importantly, the Markov approximation does not neglect
the spatial propagation phases, which are fully retained
in the dyadic Green tensor. The sharp collapse of
specular reflection across the Rayleigh threshold is
therefore a geometric diffraction effect, rather than
a consequence of non-Markovian memory. Replacing
$\omega_0$ by $\omega_0+\Delta$ shifts the diffraction
condition only by a fractional amount of order
$|\Delta|/\omega_0$, negligible compared with the
finite-size and finite-waist rounding of the threshold.

\section{Momentum-space redistribution into near-grazing diffraction channels}
\label{sec:grazing_momentum}

To identify the radiative channels responsible for the collapse of specular reflection, we calculate the scattered single-photon momentum distribution throughout the free-space light cone, without using the paraxial approximation for the outgoing photon. The incident light is the same quasi-monochromatic Gaussian mode as in Sec.~III, with atomic-site amplitudes $\psi_j$ given by Eq.~\eqref{eq:psiSM}. At each lattice constant, the incident frequency is chosen as $\omega_{\mathrm{in}}=\omega_0+\Delta_{\max}(a)$, where $\Delta_{\max}$
maximizes $R_{\mathrm{spec}}$. The single-photon response is described by the Green function $\mathcal G_{ij}(\omega_{\mathrm{in}})$ already defined in Eq.~\eqref{eq:Gij2SM}.

\begin{figure}
\includegraphics[width=12cm]{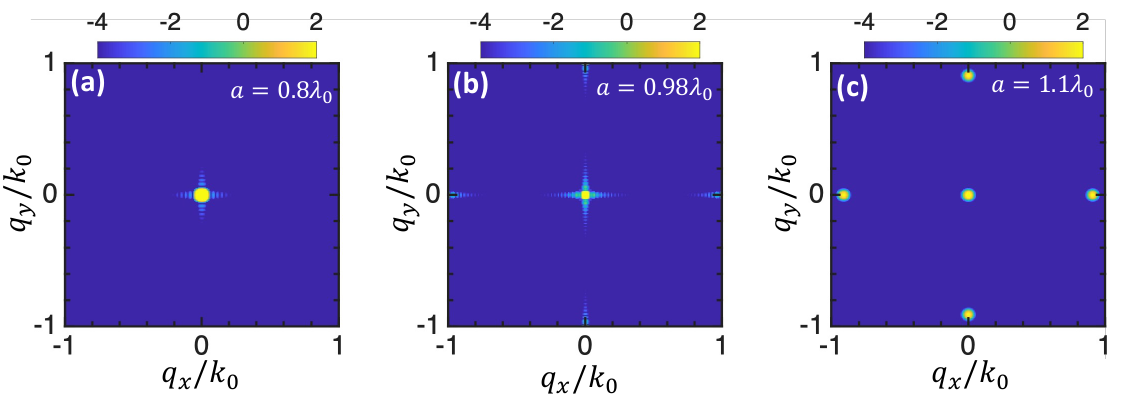}
\centering
  \caption{\textbf{Full-angle transverse-momentum distribution
    of scattered single photons across the Rayleigh threshold.}
    The color scale shows
    $\log_{10}[dP_{\mathrm{sc}}/d^2(\boldsymbol{q}/k_0)]$,
    summed over both outgoing hemispheres, for
    (a) $a=0.8\lambda_0$,
    (b) $a=0.98\lambda_0$, and
    (c) $a=1.1\lambda_0$.
    All panels use the same color scale, and the incident
    detuning is chosen as $\Delta_{\max}$ at each $a$.
    Scattering is predominantly specular below threshold;
    near threshold, intensity builds up along the four
    lattice axes close to the light-cone boundary.
    Above threshold, four first-order diffraction peaks
    appear at $|\boldsymbol{q}|/k_0\simeq\lambda_0/a$.
    The square array has $L=40$ and $w_0=10\lambda_0$.}
    \label{fig:full_angle_grazing}
\end{figure}

For an outgoing photon with $\boldsymbol{k}=(\boldsymbol{q},\pm k_z)$, where $k_z=(k_0^2-|\boldsymbol{q}|^2)^{1/2}$, the full vectorial dipole coupling is retained. For the in-plane circular dipole $\boldsymbol{e}_+$ of Eq.~\eqref{eq:gklambda}, the sum over the two transverse output polarizations is
$\sum_\lambda|\boldsymbol{e}_{\boldsymbol{k}\lambda}^*\cdot\boldsymbol{e}_+|^2=[1+(k_z/k_0)^2]/2$. Denoting by $P_{\mathrm{sc}}$ the squared norm of the scattered-field component of the outgoing one-photon state, the transverse-momentum density, summed over both outgoing
hemispheres, is
\begin{equation}
\label{eq:grazing_full_angle_density}
\begin{aligned}
\frac{dP_{\mathrm{sc}}}{d\boldsymbol{q}}
=&\frac{9\gamma_0^2}{16k_0^4}
\frac{1+(k_z/k_0)^2}{k_z/k_0}\left|\sum_{i,j}
e^{-i\boldsymbol{q}\cdot\boldsymbol{R}_i}
\mathcal G_{ij}(\omega_{\mathrm{in}})\psi_j\right|^2,
\qquad |\boldsymbol{q}|<k_0.
\end{aligned}
\end{equation}
Here $P_{\mathrm{sc}}$ is not the total outgoing-photon probability. It omits interference between the scattered field and the directly transmitted incident field. In the lossless model, the full-angle integral obeys the optical-theorem identity
\begin{equation}
\label{eq:grazing_optical_theorem}
P_{\mathrm{sc}}\equiv
\int_{|\boldsymbol{q}|<k_0}d\boldsymbol{q}\,
\frac{dP_{\mathrm{sc}}}{d\boldsymbol{q}}
=-2\operatorname{Re}(t_{\mathrm{spec}}-1)
=2R_{\mathrm{spec}}+D_{\mathrm{dif}},
\end{equation}
where $D_{\mathrm{dif}}$ is defined in Eq.~(S84). For the dimensionless axes $\boldsymbol{q}/k_0$ used in Fig.~\ref{fig:full_angle_grazing}, the plotted density is $\log_{10}dP_{\mathrm{sc}}/d(\boldsymbol{q}/k_0)$. Near the light-cone boundary, the integrable factor $1/k_z$ arises from the transformation $d\Omega=d^2\boldsymbol{q}/(k_0k_z)$ for each hemisphere. The numerical distribution is therefore averaged over finite momentum bins.

Figure~\ref{fig:full_angle_grazing} directly illustrates the redistribution of scattered light. For $a=0.8\lambda_0$, the distribution is concentrated near the specular direction. Close to the threshold, at $a=0.98\lambda_0$, pronounced scattering develops along the four lattice-axis directions near the light-cone boundary. Although the first non-specular orders of an infinite square lattice remain closed for  $a<\lambda_0$, the finite array and finite beam waist broaden transverse-momentum selection and allow near-grazing emission on the subthreshold side. At $a=1.1\lambda_0$, the four first-order diffraction peaks are resolved inside the light cone at $|\boldsymbol{q}|/k_0\simeq\lambda_0/a\simeq0.91$. Their emergence provides a direct far-field signature of the radiative-channel opening underlying the enhanced collective linewidth in Eq.~\eqref{eq:GammaqSM} and the suppression of specular reflection. These nearly grazing photons lie outside the low-NA collection cone used for the $2f$
coincidence measurements in Sec.~VI.

\bibliography{main}